%% file: main.tex
\documentclass[envcountsame,11pt]{llncs}
\usepackage{geometry}
\newenvironment{reftheorem}[2]{\begin{trivlist}
\item[\hskip \labelsep {\bfseries #1}\hskip \labelsep {\bfseries #2.}]\itshape}{\end{trivlist}}

\usepackage{latexsym}
\usepackage{tikz}
\usetikzlibrary{arrows}
\usepackage[intlimits]{amsmath}
\usepackage{txfonts}
\usepackage[ruled,vlined,linesnumbered,lined]{algorithm2e}

\newcommand{\A}{\mathcal{A}}

\newcommand{\C}{\mathcal{C}}
\newcommand{\D}{\mathcal{D}}
\newcommand{\Dz}{\mathcal{D}^{\rightarrow}}
\newcommand{\Dn}{\mathcal{D}^{\leftrightarrow}}

\newcommand{\F}{\mathcal{F}}

\newcommand{\calF}{\mathcal{F}}

\newcommand{\M}{\mathcal{M}}

\newcommand{\HR}{\mathit{HR}}
\newcommand{\MD}{\mathit{MD}}
\newcommand{\SMD}{\mathit{CMD}}

\newcommand{\V}{\mathcal{V}}
\newcommand{\calP}{\mathcal{P}}
\newcommand{\calO}{\mathcal{O}}

\newcommand{\bbtran}[1]{\stackrel{#1}{\leadsto}}

\newcommand{\Val}{\mathit{Val}}
\newcommand{\Prob}{\mathit{Prob}}
\newcommand{\Reach}{\mathit{Reach}}
\newcommand{\run}{\mathit{Run}}
\newcommand{\len}{\mathit{len}}

\newcommand{\Nset}{\mathbb{N}}
\newcommand{\Zset}{\mathbb{Z}}
\newcommand{\Qset}{\mathbb{Q}}
\newcommand{\Rset}{\mathbb{R}}
\newcommand{\Nseto}{\Nset_0}

\newcommand{\PSPACE}{\textbf{PSPACE}}
\newcommand{\NP}{\textbf{NP}}
\newcommand{\coNP}{\textbf{coNP}}

\newcommand{\ST}{\mathit{ST}}
\newcommand{\NT}{\mathit{NT}}
\newcommand{\ValOne}{\mathit{ValOne}}
\newcommand{\OptValOne}{\mathit{OptValOne}}
\newcommand{\CN}{\mathit{CN}}
\newcommand{\MP}{\mathit{MP}}

\newcommand{\SMP}[3]{\Sigma^{MP}}
\newcommand{\SCN}[1]{\Sigma^{CN}}
\newcommand{\SDN}[1]{\Sigma^{DN}}

\newcommand{\CNCO}[1]{\CN_{#1}}
\newcommand{\CNCS}[2]{\CN^{#1}_{#2}}

\newcommand{\CNDS}[3]{\CN^{#1}_{#2,#3}}
\newcommand{\CNDSNO}[3]{\CN^{#1}}
\newcommand{\CNDONO}[2]{\CN}

\newcommand{\MPS}[3]{\mathit{MP}^{#1}_{#2,#3}}
\newcommand{\MPONO}[2]{\mathit{MP}}
\newcommand{\MPSNO}[3]{\mathit{MP}^{#1}}

\newcommand{\bscc}[1]{\mathit{BSCC}[#1]}
\newcommand{\rwV}[3]{V[#1,#2]} %
\newcommand{\rwiV}[4]{V_{#1}[#2,#3]} %

\newcommand{\eps}{\varepsilon}

\newcommand{\tran}[1]{{}\mathchoice%
    {\stackrel{#1}{\rightarrow}}
    {\mathop {\smash\rightarrow}\limits^{\vrule width 0pt height 0pt depth 4pt\smash{#1}}}
    {\stackrel{#1}{\rightarrow}}
    {\stackrel{#1}{\rightarrow}}
{}}
\newcommand{\btran}[1]{{}\mathchoice%
    {\stackrel{#1}{\hookrightarrow}}
    {\mathop {\smash\hookrightarrow}\limits^{\vrule width 0pt height 0pt depth 4pt\smash{#1}}}
    {\stackrel{#1}{\hookrightarrow}}
    {\stackrel{#1}{\hookrightarrow}}
{}}
\newcommand{\ctran}[1]{{}\mathchoice%
    {\stackrel{#1}{\mapsto}}
    {\mathop {\smash\mapsto}\limits^{\vrule width 0pt height 0pt depth 4pt\smash{#1}}}
    {\stackrel{#1}{\mapsto}}
    {\stackrel{#1}{\mapsto}}
{}}

\input{nobound-statements}

\begin{document}

\title{One-Counter Markov Decision Processes}

\author{ T. Br\'{a}zdil \inst{1}
\and     V. Bro\v{z}ek\inst{1}
\and
        K. Etessami\inst{2}
\and
A. Ku\v{c}era\inst{1}
\and
D. Wojtczak\inst{2,3}
 }

\institute{
Faculty of Informatics, Masaryk University \\
\email{\{xbrazdil,xbrozek,tony\}@fi.muni.cz}
\and
School of Informatics, University of Edinburgh \\
\email{kousha@inf.ed.ac.uk} \and
CWI, Amsterdam \\
\email{D.K.Wojtczak@cwi.nl}}

\maketitle
{\raggedbottom

\begin{abstract}

\normalsize\input{abstract}

\end{abstract}
\pagebreak
\setcounter{page}{1}
}

\vspace*{-0.1in}
\input{intro}

\vspace*{-0.1in}

\input{defs}

\vspace*{-0.1in}

\input{nobound}

\vspace*{-0.15in}

\input{bound}

\newpage
\appendix

\input{app-nobound}
\input{app-bound}

\section{Why bounding the counter can yield bad approximations}
\label{app-easy-qbd}

As discussed in the introduction, here is a simple
example for why cutting off the counter at a finite value,
even for a purely stochastic QBD
(equivalently, a probabilistic one-counter automaton)
can in general radically alter its behavior.
Consider a 2-state QBD which in state 1,
with probability $p = 1/2^n$ goes to state 2, and with probability
$1-p$ stays in state $1$, in both cases incrementing the counter, and
in state 2 stays in state 2 with probability 1 and decrements the counter.
We are interested in the probability of termination starting at state 1,
with counter value 1.
By cutting off the counter at a value $N \in 2^{o(n)}$
the termination probability goes down to $\epsilon$ arbitrarily close to $0$,
for large enough $n$.
Although we used small probabilities $1/2^n$ in this example,
the same thing can easily be achieved using a QBD with $O(n)$ states
and only the probability $1/2$ on transitions.

\end{document}

%% file: nobound-statements.tex
\newcommand{\sbLEOM}{
Given a OC-MDP, $\A$, there
is a finite-state MDP  with rewards, $\D$, computable
in polynomial time from $\A$,
such that the set of vertices of $\D$ contains $Q$ and
for every $p\in Q$, $i\in \Zset$ we have that
$\Val^{\CNCO{\A}}(p(i))=\Val^{\CNDONO{\D}{r}}(p)$.
Moreover,
for a MD strategy $\sigma$ in $\D$, let $\sigma'$ be
the 
CMD strategy in $\Dn_\A$ with a selector $f$ %
defined by $f(p)=\sigma(p)$.
Then for each $p(i)\in Q\times \Zset$, 
$\calP(\CNCS{\sigma'}{\A}(p(i)))=\calP(\CNDSNO{\sigma}{\D}{r}(p))$.
}

\newcommand{\sbPROOR}{
Let
$A \coloneqq \{ v\in V \mid \Val^{\CNDONO{\D}{r}}(v)=1\}$.
Then for all $u\in V$ we have:
\vspace*{-0.1in}
\[
\Val^{\CNDONO{\D}{r}}(u) = 
\mathop{\max\vphantom{p}}\limits_{\tau\in {MD}}\calP({Reach}^{\tau}_{A}(u))=
\sup_{\tau\in {HR}}\calP({Reach}^{\tau}_{A}(u))
\]
}

\newcommand{\sbLED}{
$\SCN{\D}$ is the set of all decreasing strategies from $\SMP{\D}{r}{\leq 0}$.
}

%% file: abstract.tex
We study the computational complexity of central analysis problems for
One-Counter Markov Decision Processes (OC-MDPs), a class of
finitely-presented, countable-state MDPs.

OC-MDPs extend finite-state MDPs with an unbounded
counter.  The counter can be incremented, decremented, or not changed 
during each
state transition, and transitions may be enabled or not depending on
both the current state and on whether the counter value is 0 or not.
Some states are ``random'', from where the next transition is chosen
according to a given probability distribution, while other states are
``controlled'', from where the next transition is chosen by the
controller.  Different objectives for the controller give rise to
different computational problems, aimed at computing optimal achievable
objective values and optimal strategies.

OC-MDPs are in fact equivalent to a controlled extension of
(discrete-time) Quasi-Birth-Death processes (QBDs), a purely
stochastic model heavily studied in queueing theory and applied
probability.  They can thus be viewed as a natural ``adversarial''
extension of a classic stochastic model.  They can also be
viewed as a natural probabilistic/controlled extension of classic
one-counter automata.  OC-MDPs also subsume (as a very restricted
special case) a recently studied MDP model called ``solvency games''
that model a risk-averse gambling scenario. 

Basic computational questions for OC-MDPs include ``termination''
questions and ``limit'' questions, such as the following: does the
controller have a strategy  to ensure that the
counter (which may, for example, count the number of jobs in the
queue) will hit value 0 (the empty queue) almost surely (a.s.)?  Or
that the counter will have $\lim \sup$ value $\infty$, a.s.?  Or, that
it will hit value 0 in a selected terminal state, a.s.?  Or, in case
such properties are not satisfied almost surely, compute their optimal
probability over all strategies.

We provide new upper and lower bounds on the complexity of such
problems.  Specifically, we show that several quantitative and
almost-sure limit problems can be answered in polynomial time, and
that almost-sure termination problems (without selection of desired
terminal states) can also be answered in polynomial time.  On the
other hand, we show that the almost-sure termination problem with
selected terminal states is PSPACE-hard and we provide an exponential
time algorithm for this problem.  We also characterize classes of
strategies that suffice for optimality in several of these settings.

Our upper bounds combine a number of techniques from the theory of
MDP reward models, the theory of random walks, 
and a variety of automata-theoretic methods.

%% file: intro.tex
\section{Introduction}
\label{sec-intro}

Markov Decision Processes (MDPs) are a standard model for
stochastic dynamic optimization.  
They describe a system that exhibits both stochastic and controlled
behavior. The system begins in some state
and makes a sequence of state transitions;
depending on the state,
either the controller gets to 
choose from among possible transitions, or there is a probability
distribution over possible transitions.\footnote{Our focus is 
on discrete state spaces, and discrete-time MDPs.  In some presentations
of such MDPs, probabilistic and controlled transitions are combined into one: each transition entails
a controller move followed by a probabilistic move.   The two presentations
are equivalent.}
Fixing a {\em strategy} for the controller
determines a probability space of
(potentially infinite) runs, or trajectories, of the MDP.
The controller's goal is to
optimize the (expected) value of some objective function,
which may be a function of the entire trajectory.
Two fundamental computational questions that arise are
``{\em what is the optimal value that the controller can achieve?}''
and ``{\em what strategies achieve this?}''. 
For finite-state MDPs, such questions have 
been studied
for many objectives and
there is a large literature on both the  
complexity of central questions
as well as on methods that work well in practice,
such as value iteration and policy iteration (see, e.g., \cite{Puterman94}).

Many important stochastic models are, however, not finite-state, but are
finitely-presented and describe an infinite-state underlying
stochastic process.  Classic examples include branching processes,
birth-death processes, and many others.  Computational questions for
such purely stochastic models have also been studied for a long time.  
A 
model that is of direct relevance to this paper is the 
Quasi-Birth-Death process (QBD), a generalization of birth-death
processes that has been heavily studied in queueing theory and applied
probability (see, e.g., the books
\cite{Neuts81,LatRam99,BiLaMe05,GrossHarris98}).
Intuitively, a QBD describes an unbounded queue, using a counter to count
the number of jobs in the queue, and such that the queue can be in one of a
bounded number of distinct ``modes'' or ``states''.  
Stochastic transitions can add or remove jobs from the queue and
can also transition the queue from one state to another.
QBDs are in general studied as continuous-time processes,
but many of their key analyses (including both steady-state and
transient analyses) amount to analysis of their underlying embedded
discrete-time QBD (see, e.g., \cite{LatRam99}).  
An equivalent way to view discrete-time QBDs is as a 
probabilistic extension 
of classic {\em one-counter automata}
(see, e.g, 
\cite{ValPat73}), which extend finite-state automata 
with an unbounded
counter.
 The counter can be incremented, decremented, or remain
unchanged during 
state transitions, and transitions may be enabled or not depending on
both the current state and on whether the counter value is 0 or not.
In {\em probabilistic} one-counter automata (i.e., QBDs), 
from every state the next transition is chosen
according to a probability distribution depending on that
state.
(See \cite{EWY08} for 
more information on the relation between QBDs and other models.)

In this paper we study {\em One-Counter Markov Decision Processes} (OC-MDPs),
which extend  discrete-time QBDs with a controller.
An OC-MDP has a finite set of states:
some states are {\em random}, from where the next transition is chosen
according to a given probability distribution, and other states are
{\em controlled}, from where the next transition is chosen by the
controller.  Again, transitions can change the state and can also change the value of the (unbounded) counter by at most 1.
Different objectives for the
controller give rise to
different computational problems for OC-MDPs, aimed at optimizing 
those objectives.

Motivation for studying OC-MDPs comes from several different
directions.  Firstly, 
it is very natural, both in queueing theory and in other contexts, to
consider an ``adversarial'' extension of stochastic models like QBDs,
so that
stochastic assumptions can sometimes be replaced by ``worst-case'' 
or ``best-case''
assumptions. 
For example, under stochastic assumptions about arrivals, we
may wish to know whether there exists a ``best-case'' control of the
queue under which the queue will almost surely become empty (such
questions are of course related to the stability of the queue), 
or we
may ask if we can do this with at least a given probability.  Such
questions are similar in spirit to questions asked in the
rich literature on ``adversarial queueing theory'' (see, e.g.,
\cite{BKRSW01}), although this is a somewhat different setting.  These
considerations lead naturally 
to the extension
of QBDs with control, and thus to OC-MDPs. 
Indeed, MDP variants of QBDs have already been studied in 
the stochastic modeling literature, see \cite{White05,LHB07}.
However, in order to keep their analyses tractable, these works take
the drastic approach of cutting off the value of the counter 
(i.e., size of the queue) at
some arbitrary finite value $N$, effectively adding dead-end absorbing states
at values higher than $N$.
This restricts the model to
a finite-state ``approximation''.
However, cutting off the counter value can in fact radically alter the
behavior of the model,
even for purely probabilistic QBDs
(see appendix \ref{app-easy-qbd} for simple examples).
Thus the existing work in the QBD literature on MDPs does not
establish any results about the computational complexity,
or even decidability, of basic analysis problems for general OC-MDPs.

OC-MDPs also subsume another recently studied infinite-state MDP model
called {\em solvency games} \cite{BKSV08},
which amount to a very limited subclass of OC-MDPs. 
Solvency games model
a risk-averse ``gambler'' (or ``investor'').
The gambler has an initial pot of money, given by
a positive integer, $n$.
He/she then has to
choose repeatedly from among a finite set of possible gambles, 
each of
which has an associated random gain/loss given by a finite-support 
probability distribution over the integers.
Berger et. al. \cite{BKSV08} study 
the gambler objective of   
minimizing the probability of going bankrupt.
One can of course study the same basic repeated
gambling model under a variety of other objectives, and many such objectives
have been studied.     
It is not hard to see that all such repeated gambling models constitute  
special cases of 
OC-MDPs.
The counter in an OC-MDP can keep track of the gambler's wealth.
Although, by definition, OC-MDPs can only increment or decrement
the counter by one in each state transition, it is easy to 
augment any finite change to the counter value by using auxiliary
states and incrementing or decrementing the counter by one at a time.  
Similarly, with an OC-MDP
one can easily augment any choice over finite-support
probability distribution on integers, each of which defines the random 
change to the counter
corresponding to a particular gamble. 
\cite{BKSV08} showed that 
if the solvency game satisfies 
several additional restrictive technical
conditions, 
then one can characterize 
the optimal strategies for minimizing the probability
of bankruptcy (as a kind of ``ultimately memoryless'' strategy)
and compute them using linear programming.
They did not however establish any results for general, unrestricted, solvency 
games.  They conclude with the following remark:
{``It is clear that our results are at best a sketch of basic
elements of a larger theory''}.   We believe OC-MDPs constitute 
an appropriate larger framework
within which to study algorithmic questions not just for solvency games,
but for various more general infinite-state MDP models
that employ a counter.
In Section \ref{sec-boundary}, Proposition \ref{prop-solvency}, 
we show that all {\em qualitative}
questions about (unrestricted) solvency games, namely whether the gambler
has a strategy to not go bankrupt with probability $>0$, $=1$, $=0$, $<1$, can be answered in polynomial time.

Our goal it to study the computational complexity of 
central analysis problems
for OC-MDPs.  
Key quantities
associated with discrete-time QBDs, which can be used to derive many
other useful quantities, are ``termination probabilities'' (also known as
their ``$G$ matrix'').  These are the probabilities that,
starting from a given state, with counter value 1, we will
eventually reach counter value $0$ for the first time in some other
given state.  The complexity of computing termination probabilities
for QBDs is already an intriguing problem, and many 
numerical methods have been devised for it.  A recent result
in \cite{EWY08} shows that these probabilities can be approximated in
time polynomial in the size of the QBD, in the unit-cost RAM model of
computation, using a variant of Newton's method,  
but that deciding , e.g., whether a termination
probability is $\geq p$ for a given rational $p \in (0,1)$ in the
standard Turing model is at least as hard as a long standing open
problem in exact numerical computation, namely the square-root sum
problem, which is not even known to be in NP nor the polynomial-time
hierarchy. (See \cite{EWY08} for
more information.)

We study OC-MDPs under related objectives,
in particular, 
the objective of maximizing termination probability,
and of maximizing the probability of termination in a particular subset of the states 
(the latter problem is considerably harder, as we shall see).
Partly as a stepping stone 
toward these objectives, but also for its own intrinsic interest, we also    
consider OC-MDPs without boundary, meaning where the counter can take 
on both positive and negative
values, and we study the objective of optimizing the probability that the 
$\lim \sup$ value is $= \infty$ (or, by symmetry, that the $\lim \inf$
is $= -\infty$).
The boundaryless model is related, in a rather subtle way, 
to the well-studied model of finite-state MDPs with limiting average reward 
objectives (see, e.g., \cite{Puterman94}).
This connection enables us to exploit 
recent results for finite-state MDPs (\cite{Gimbert-STACS07}), and classic
facts in the theory of 1-dimensional random walks and sums of i.i.d. random variables, to analyze the 
boundaryless case of OC-MDPs.
We then use these analyses as crucial building blocks for the analysis of optimal termination probabilities
in the case of OC-MDPs with boundary.
Our main results are the following:

\begin{enumerate}

\item For boundaryless OC-MDPs, where the objective of the controller
is to maximize the probability that the $\lim \sup$ ($\lim \inf$) of the counter value
in the run (the trajectory) is $\infty$ ($-\infty$), the situation is
as good as we could hope. Namely, we show:
 \begin{enumerate}
   \item  The optimal probability is a rational value that is polynomial-time 
computable.
 
   \item  There exist deterministic optimal strategies
          that are both ``{\em counter-oblivious}''
          and {\em memoryless} (we shall call these CMD strategies),
          meaning the choice of the next transition depends only on the current state
          and neither on the history, nor on the current counter value.

          Furthermore, such an optimal strategy can be computed in polynomial time.

 \end{enumerate}

\item For OC-MDPs with boundary, where the objective is to maximize the probability
      that, starting in some state
      and with counter value 1, we eventually {\em terminate} (reach counter value $0$) {\em in any state},
      we have:

     \begin{enumerate}
      \item  In general the optimal (supremum) probability can be an irrational value, and this is so
              already in the case of QBDs where there is no controller, 
              see \cite{EWY08}.

       \item It is decidable in polynomial time whether the optimal probability is 1.

       \item There is a CMD strategy such that starting from every state with value 1,
             using that strategy we terminate almost surely.

             (Optimal CMD strategies need not exist starting from states where the optimal probability is not 1.) 
      \end{enumerate}

\item For OC-MDPs with boundary, where the objective is to maximize the probability that, starting from
      a given state and counter value 1, we terminate in a {\em selected} subset of states $F$ (i.e., 
      reach counter value $0$ for the first time in one of these selected states), we know the following:

     \begin{enumerate}
      \item  The optimal probabilities can of course again be irrational.

      \item  There need not exist any optimal strategy, even when the supremum probability of termination
             in selected states is 1 (i.e., only $\epsilon$-optimal strategies may exist).

       \item  Even deciding whether there is an optimal strategy which ensures probability 1
              termination in the selected states is  PSPACE-hard.%

       \item  We provide an exponential time algorithm  to determine whether there is a strategy
              using which  the probability of termination in the
selected states is 1, starting at a given 
state and counter value.
 
     \end{enumerate}
\end{enumerate}

Our proofs employ techniques from several areas:
from the theory of finite-state MDP reward models (including some recent results), 
from the theory of 1-dimensional random walks
and sums of i.i.d. random variables, and  a variety of
automata-theoretic methods (e.g., pumping arguments, 
decomposition arguments, etc.). 
Our results leave open many fascinating questions about OC-MDPs.  
For example, we do not know 
whether the following problem is decidable:  given an OC-MDP and a rational probability $p \in (0,1)$,  
decide whether the optimal probability of
termination (in any state) is $> p$.
Other open questions pertain to OC-MDPs where the objective is to minimize termination probabilities.
We view this paper as laying the basic foundations for the algorithmic analysis of OC-MDPs,
and we feel that answering some of the remaining open questions will 
likely reveal an even
richer underlying theory.

{\bf Related work.}    
A more general MDP model that strictly subsumes OC-MDPs, called {\em Recursive Markov Decision Processes} (RMDPs) 
was studied in \cite{EY05icalp,EY06stacs}.   
These are equivalent to MDPs whose state transition structure
is that of a general pushdown automaton.
Problems such as deciding whether there is a strategy that
yields termination probability 1,
or even approximating the maximum probability within any non-trivial additive factor, were shown 
to be undecidable for general RMDPs in \cite{EY05icalp}.
For the restricted class of 1-exit RMDPs  (which correspond in a precise sense to MDP versions
of multi-type branching processes, stochastic context-free
grammars,   and a related model called pBPAs), \cite{EY05icalp}
showed quantitative problems for optimal termination 
probability are decidable in PSPACE, and
\cite{EY06stacs} showed that deciding whether the optimal termination
probability is 1  can be done in P-time.
In \cite{BBFK06} this was extended further to answer qualitative almost-sure 
reachability questions for 1-exit RMDPs in P-time. 
1-exit RMDPs are however incompatible with OC-MDPs (which
actually correspond to 1-box RMDPs). 
The references in these cited papers point to earlier related literature, in particular on probabilistic Pushdown Systems 
and Recursive Markov chains.
There is a substantial literature on numerical algorithms for analysis
of QBDs and related purely stochastic models (see
\cite{Neuts81,LatRam99,BiLaMe05}).  In that literature one can 
find results related to qualitative questions, like whether the
termination probability for a given QBD is 1.  Specifically, it is
known that for an {\em irreducible} QBD, i.e., a QBD in which from
every configuration (counter value and state) one can reach
every other configuration with non-zero probability, 
whether the underlying Markov chain is recurrent
boils
down to steady-state analysis of induced finite-state chains over
states of the QBD, and in particular on whether the expected
one-step change in the counter value in steady state is $\leq 0$ (see,
e.g., Chapter 7 of \cite{LatRam99} for a proof).
However, these results crucially assume the QBD
is irreducible.  They do not directly yield an algorithm for deciding, for 
general QBDs, whether 
the probability of termination is 1
starting from a given state and counter value 1.  Thus, our results
for OC-MDPs yield new
results even for purely stochastic QBDs without 
controller.

%% file: defs.tex
\section{Basic definitions}
\label{sec-defs}

\noindent
We use $\Zset$, $\Nset$, $\Nset_0$, 
to denote the integers, positive integers, and non-negative 
integers, respectively. We use 
standard notation for intervals, e.g.,
$(0,1]$ denotes $\{x \in \Rset \mid 0 < x \leq 1 \}$. 
The set of finite words over an alphabet $\Sigma$ is denoted
$\Sigma^*$, and the set of infinite words over $\Sigma$ is denoted
$\Sigma^\omega$.  $\Sigma^+$ 
denotes $\Sigma^* \smallsetminus \{\varepsilon\}$ where 
$\varepsilon$ is the empty word. The length of a given $w \in \Sigma^* \cup
\Sigma^{\omega}$ is denoted $\len(w)$, where the length of an infinite 
word is $\infty$. 
Given a word (finite or infinite) over $\Sigma$, the individual
letters of $w$ are denoted $w(0),w(1),\cdots$  (so indexing begins at $0$). 
For a word
$w$, we denote by $w{\downarrow} n$ the prefix $w(0)\cdots w(n{-}1)$
of $w$.
Let $\V = (V,\tran{})$ where $V$ is a non-empty set and 
${\tran{}} \subseteq V \times V$ a \emph{total} relation 
(i.e., for every $v \in V$ there is some $u \in
V$ such that $v \tran{} u$). The reflexive transitive closure of
$\tran{}$ is denoted $\tran{}^*$. A \emph{path} in $\V$ is a finite 
or infinite word $w \in V^+ \cup V^\omega$ such that 
$w(i{-}1) \tran{} w(i)$ for every $1 \leq i < \len(w)$.
A \emph{run} in $\V$ is an infinite path in $V$. The set of all runs
in $\V$ is denoted $\run_{\V}$. The set of runs in $\V$
that start with a given finite path $w$ is denoted $\run_{\V}(w)$.

We assume familiarity with basic notions of probability, e.g.,
a $\sigma$-field,  $\calF$, over a set $\Omega$, and
a probability measure $\calP: \calF \mapsto [0,1]$, together define a
{\em probability space}  $(\Omega,\calF,\calP)$.
As usual, a \emph{probability distribution} over a finite or 
countably infinite 
set $X$ is a function
$f : X \rightarrow [0,1]$ such that \mbox{$\sum_{x \in X} f(x) = 1$}. 
We call $f$ \emph{positive} if 
$f(x) > 0$ for every $x \in X$, and \emph{rational} if $f(x) \in
\Qset$ for every $x \in X$.

For our purposes, a \emph{Markov chain} is  a
triple \mbox{$\M = (S,\tran{},\Prob)$}
  where $S$ is a finite or countably infinite set of \emph{states},
  ${\tran{}} \subseteq S \times S$ is a total \emph{transition relation},
  and $\Prob$ is a function that assigns to each state $s \in S$ 
  a positive probability distribution over the outgoing transitions
  of~$s$. As usual, we write $s \tran{x} t$ when $s \tran{} t$
  and $x$ is the probability of $s \tran{} t$. 
To every $s \in S$ we associate the probability 
space $(\run_{\M}(s),\calF,\calP)$ of runs starting at $s$,
where 
$\calF$ is the \mbox{$\sigma$-field} generated by all \emph{basic cylinders},
$\run_{\M}(w)$, where $w$ is a finite path starting with~$s$, and 
$\calP: \calF \rightarrow [0,1]$ is the unique probability measure such that
$\calP(\run_{\M}(w)) = \prod_{i{=}1}^{\len(w)-1} x_i$ where
$w(i{-}1) \tran{x_i} w(i)$ for every $1 \leq i < \len(w)$.
If $\len(w) = 1$, we put $\calP(\run_{\M}(w)) = 1$. 

\begin{definition} 
A {\bf Markov decision process (MDP)} is a tuple 
$\D = (V,{\btran{}},(V_N,V_P),\Prob)$, where $V$ is a finite or
countable set of \emph{vertices}, ${\btran{}} \subseteq V \times V$ 
is a total \emph{transition relation}, $(V_N,V_P)$ is a
partition of $V$ into \emph{non-deterministic} (or ``controlled'') and \emph{probabilistic}
vertices, and $\Prob$ is a \emph{probability assignment} which
to each $v \in V_P$ assigns a rational probability distribution on
its set of outgoing transitions. 
\end{definition}

\noindent
A \emph{strategy} is a function
$\sigma$ which to each $wv \in  V^*V_N$ assigns a
probability distribution on the set of outgoing transitions of~$v$.
We say that a strategy
$\sigma$ is \emph{memoryless (M)} if $\sigma(wv)$ depends only on the
last vertex $v$, and \emph{deterministic (D)} if $\sigma(wv)$ is a
Dirac distribution (assigns probability 1 to some transition) for each $wv \in V^*V_{N}$. 
When $\sigma$ is D, we write $\sigma(wv)=v'$ instead
of $\sigma(wv)(v,v')=1$. For a MD strategy $\sigma$, we write 
$\sigma(v) = v'$ instead of $\sigma(wv)(v,v')=1$.
Strategies that are 
not necessarily
memoryless (respectively, deterministic) are called 
\emph{history-dependent (H)} (respectively,
 \emph{randomized (R)}). 
We use $\HR$ to denote the set of all (i.e., $H$ and $R$) strategies, 
and we use similar suggestive notation for other strategy classes.

Each strategy $\sigma$ determines a unique Markov chain $\D(\sigma)$
for which $V^+$ is the set of states, and $wu \tran{x} wuu'$ iff 
$u \btran{} u'$ {\em and} one of the following conditions holds:
(1) $u \in V_{P}$ and $\Prob(u,u') = x$, or
(2) $u \in V_{N}$ and $\sigma(wu)$ assigns $x$ to the transition $(u,u')$. 
To every $w \in \run_{\D(\sigma)}$ we associate the corresponding run
$w_{\D} \in \run_{\D}$ where $w_{\D}(i)$ is the vertex currently \emph{visited}
by $w(i)$, i.e., the last element of $w(i)$ (note $w(i) \in V^+$).

For our purposes in this paper, an \emph{objective}\footnote{In general, objectives
can be arbitrary Borel measurable functions of trajectories, for
which we want to optimize expected value.  We only consider
objectives that are characteristic functions of a measurable set
of trajectories.}
is a set\ 
 $O \subseteq \run_{\D}$ (in situations when 
the underlying MDP $\D$ is not clear from the context, 
we write $O_\D$ instead of $O$). For every strategy
$\sigma$, let $O^\sigma$ be the set of all $w \in \run_{\D(\sigma)}$ such that
$w_\D \in O$. Further, for every $v \in V$ we use $O^\sigma(v)$ to denote 
the set of all $w \in O^\sigma$ which start at~$v$. We say that $O$ is
\emph{measurable} if $O^\sigma(v)$ is measurable for all $\sigma$ and
$v$. For a measurable objective $O$ and a vertex $v$, the 
\emph{$O$-value in $v$} is defined as follows:
$\Val^{O}(v)  =  \sup_{\sigma \in \HR} \calP(O^\sigma(v))
$.
We say that a strategy $\sigma$ is \emph{$O$-optimal} starting at 
a given vertex $v$
if $\calP(O^\sigma(v)) = \Val^{O}(v)$. We say $\sigma$ is {\em $O$-optimal},
if it is optimal starting at every vertex.
An important objective for us is 
\emph{reachability}. For every set
$T \subseteq V$ of \emph{target vertices}, we define the objective
$\Reach_T = \{w \in \run_{\D} \mid \exists i \in \Nset_0 \; s.t. \; w(i) \in T \}$.

\begin{definition}
\label{def-pPDA}  
A {\bf one-counter MDP (OC-MDP)} is a tuple,
$\A = (Q,\delta^{=0},\delta^{>0},(Q_N,Q_P),P^{=0},P^{>0})$,
where 
\begin{itemize}
  \item $Q$ is a finite set of \emph{states}, partitioned into {\em non-deterministic}, $Q_N$, and {\em probabilistic}, $Q_P$, states. 
  \item $\delta^{>0} \subseteq Q \times \{-1,0,1\} \times Q$ and 
    $\delta^{=0} \subseteq Q \times \{0,1\} \times Q$ are the sets 
    of \emph{positive} and \emph{zero rules} (transitions) such that each 
    $p \in Q$ has an outgoing positive rule and an outgoing zero rule;
  \item $P^{>0}$ and $P^{=0}$ are \emph{probability assignments}: 
    both assign 
    to each $p \in Q_P$, a positive rational probability
    distribution over the outgoing 
    transitions in $\delta^{>0}$ and $\delta^{=0}$,
    respectively, of $p$.
\end{itemize}

\end{definition}

\noindent
Each OC-MDP, $\A$, 
naturally determines an infinite-state MDP with or without a boundary,
depending on whether zero testing is taken into account or not.
Formally, we define MDPs $\Dz_{\A}$ and $\Dn_{\A}$ as follows:
\begin{itemize}
\item $\Dz_\A=(Q \times \Nseto,{\ctran{}}, 
   (Q_N \times \Nseto,Q_P \times \Nseto), \Prob)$. %
   Here for all $p,q \in Q$ and $j \in \Nseto$ we have that 
      $p(0) \ctran{} q(j)$ iff $(p,j,q) \in \delta^{=0}$. If $p \in Q_P$,
       then the probability of $p(0) \ctran{} q(j)$ is $P^{=0}(p,j,q)$. %
   Further for all $p,q \in Q$, $i \in \Nset$, and $j \in \Nseto$ we 
      have that $p(i) \ctran{} q(j)$ iff $(p,j{-}i,q) \in \delta^{>0}$.
      If $p \in Q_P$,
      then the probability of $p(i) \ctran{} q(j)$ is $P^{>0}(p,j{-}i,q)$.
\item $\Dn_\A=(Q \times \Zset,{\ctran{}}, 
      (Q_N \times \Zset,Q_P \times \Zset), \Prob)$, where 
   for all $p,q \in Q$ and $i,j \in \Zset$ we 
      have that $p(i) \ctran{} q(j)$ iff $(p,j{-}i,q) \in \delta^{>0}$.
      If $p \in Q_P$,
      then the probability of $p(i) \ctran{} q(j)$ is $P^{>0}(p,j{-}i,q)$.
\end{itemize}

\noindent
Since the MDPs $\Dz_\A$ and $\Dn_\A$ have infinitely many vertices,
even $\MD$ strategies are not necessarily finitely representable.
But the objectives we consider are often
achievable with strategies that use only
finite information about the counter or even
ignore the counter value. We call a strategy, $\sigma$, 
in $\Dz_\A$ or $\Dn_\A$, \emph{counter-oblivious}-MD (denoted $\SMD$) 
if there is a {\em selector}, 
$f : Q \tran{} \delta^{>0}$ (which selects a transition out of each state)
so that at any configuration $p(n)\in Q\times \Nset$, 
$\sigma$  
chooses transition $f(p)$ with prob. 1  
(ignoring history and $n$).

%% file: nobound.tex
\section{OC-MDPs Without Boundary}\label{sec-noboundary}
In this section we study the objective
``Cover Negative'' (CN), which says that values of the counter
during the run should cover arbitrarily low negative numbers
in $\Zset$ (i.e., that the $\lim \inf$
counter value is $= -\infty$).
Our goal is to prove Theorem~\ref{thm:optimal}.
\ (All proofs missing in this section can be found
in the Appendix.)

\begin{definition}
Let $\A$ be a OC-MDP.
We use $\CNCO{\A}$ to denote the set of all runs $w\in \run_{\Dn_\A}$ 
such that
for every $n\in \mathbb{Z}$ the run $w$ visits a configuration
$p(i)$ for some $p\in Q$ and $i\leq n$. 
\end{definition}

\begin{theorem}\label{thm:optimal}
Given a OC-MDP, $\A$, there is a $\CNCO{\A}$-optimal CMD strategy 
for it, which is computable in polynomial time.  Moreover,
$\Val^{\CNCO{\A}}$ 
is rational and
computable in polynomial time.
\end{theorem}

\noindent We prove this via a sequence of reductions
to problems for finite-state MDPs with and without {\em rewards}. 
For us an \emph{MDP with reward} is equipped with 
$r:V\to \{-1,0,1\}$.
For $v=v_0 \cdots v_n\in V^+$, let $r(v) \coloneqq \sum_{i=0}^n r(v_i)$. 

\begin{definition}
We denote by $\CNDONO{\D}{r}$ the set of all $w\in \run_{\D}$ 
satisfying $\liminf_{n\to \infty} r(w{\downarrow} n)=-\infty$.
We further denote by $\MPONO{\D}{r}$
the set of all runs $w\in \run_{\D}$ 
such that 
$\lim_{n\to \infty} \frac{r(w{\downarrow}n)}{n}$ exists and
$\lim_{n\to \infty} \frac{r(w{\downarrow}n)}{n}\leq 0$.%
\footnote{``MP'' stands for ``(non-positive) Mean Payoff''.}
\end{definition}

\noindent A theorem by Gimbert (\cite[Theorem 1]{Gimbert-STACS07})  
implies there is always a $\CNDONO{\D}{r}$-optimal MD  
strategy for finite MDPs,
because (the characteristic function of) objective $\CNDONO{\D}{r}$ 
is {\em prefix-independent} and {\em submixing}
(see~Section~\ref{proof:prop:CN_optimal_RewMDP}).
Lemma~\ref{lem:equiv_OCDP_MDP} shows 
for OC-MPDs there is
always a $\CNCO{\A}$-optimal  CMD strategy.
We define several problems:

\begin{description}
\item[OC-MDP-CN:] \mbox{}\\
{\bf Input:} OC-MDP, $\A$, and $z \in \Zset$.\\
{\bf Output:} a $\CNCO{\A}$-optimal CMD strategy 
for $\A$, and $\Val^{\CNCO{\A}}(p(z))$, for every
$p\in Q$.

\item[MDP-CN:] \mbox{}\\
{\bf Input:} finite-state MDP, $\D$,  with reward function $r$.\\
{\bf Output:} a $\CNDONO{\D}{r}$-optimal MD strategy for
$\D$, and $\Val^{\CNDONO{\D}{r}}(v)$, for every vertex $v$ of $\D$.

\item[MDP-CN-qual:]\mbox{}\\
{\bf Input:} finite-state MDP, $\D$, with reward function $r$.\\
{\bf Output:}
set $A = \{ v  \mid \Val^{\CNDONO{\D}{r}}(v)=1 \}$,
and a MD strategy $\sigma$ which is $\CNDONO{\D}{r}$-optimal starting at
every $v\in A$.

\item[MDP-MP-qual:] \mbox{}\\
{\bf Input:} finite-state MDP, $\D$, with reward function $r$.\\
{\bf Output:}
set $A = \{ v  \mid \exists\sigma_v\in MD:\calP({\MPSNO{\sigma_v}{\D}{r}}(v))=1\}$,
a $\bar\sigma \in MD$ such that $\forall v \in A:\calP({\MPSNO{\bar\sigma}{\D}{r}}(v))=1$.%
\footnote{
The existence of strategy $\bar\sigma$
is a consequence of the correctness proof in Section~\ref{subsec:corr-MP}.
}
\end{description}

\vspace*{-0.1in}

\begin{proposition}\label{prop:CN-opt-comput} 
\begin{enumerate}

\item\label{enum:CN-opt-comput-red} There exist 
the following polynomial-time (Turing) reductions: 
$$
\textbf{OC-MDP-CN} \leq_P \textbf{MDP-CN} \leq_P \textbf{MDP-CN-qual} \leq_P \textbf{MDP-MP-qual}
$$

\item\label{enum:CN-opt-comput-solve}
The problem \textbf{MDP-MP-qual} can be solved in polynomial time.

\end{enumerate}
\end{proposition}

The following lemma establishes both the first reduction of
Proposition~\ref{prop:CN-opt-comput}, part \ref{enum:CN-opt-comput-red},
and the existence of $\CNCO{\A}$-optimal CMD strategies for
OC-MDPs.

\begin{lemma}\label{lem:equiv_OCDP_MDP}
\sbLEOM
\end{lemma}

Dealing with MD strategies
simplifies notation.
Although the Markov chain $\D(\sigma)$ has
infinitely many states, for a finite MDP $\D = 
(V,\btran{},(V_N,V_P),\Prob)$ and a MD strategy $\sigma$
we can replace $\D(\sigma)$ with a finite-state Markov chain 
$\D\langle\sigma\rangle$
where $V$ is the set of states, and $u \tran{x} u'$ iff 
$u \tran{x} uu'$ in $\D(\sigma)$.
This only changes notation since
for every $u\in V$ there is an isomorphism
between the probability spaces $\run_{\D(\sigma)}(u)$ and $\run_{\D\langle\sigma\rangle}(u)$
given by the bijection of runs
which maps run $w$ to $w_{\D}$, see the definition
of $\D(\sigma)$ 
in Sect. ~\ref{sec-defs}.

To finish the proof of Theorem~\ref{thm:optimal} we have to provide
the last two reductions from 
Proposition~\ref{prop:CN-opt-comput}, part \ref{enum:CN-opt-comput-red},
prove that $\Val^{\CNDONO{\D}{r}}$ is always rational,
and prove
Proposition~\ref{prop:CN-opt-comput}, part \ref{enum:CN-opt-comput-solve}.
We do these in separate subsections.

\vspace*{-0.1in}

\subsection{Reduction to Qualitative $\CNDONO{\D}{r}$}\label{subsec:CN-to-qual}
\begin{procedure}[t]
\caption{Solve-CN($\D$,$r$)}\label{proc:solve-CN}
\dontprintsemicolon
\KwData{A MDP $\D$ with reward $r$.}
\KwResult{Compute the vector $\left(\Val^{\CNDONO{\D}{r}}(v)\right)_{v\in V}$, and a $\CNDONO{\D}{r}$-optimal MD strategy, $\sigma$.}

$\left(A, \tau\right)$ $\leftarrow$ \FuncSty{Qual-CN}($\D$,$r$) \;
\nllabel{proc:solve-CN-qual-sub}

$\left(\sigma_R,\left(val_v\right)_{v\in V}\right)$ $\leftarrow$ \FuncSty{Max-Reach}($\D$,$A$) \;
\nllabel{proc:solve-CN-max-reach}

\lFor{every $v \in V_N$}%
{
\lIf{$v \in A$}{$\sigma(v) \leftarrow \tau(v)$} \lElse{$\sigma(v) \leftarrow \sigma_R(v)$} \;
\nllabel{proc:solve-CN-comp-sig}
}

\Return{$\left(val_v\right)_{v\in V}$, $\sigma$} \;
\end{procedure}

\begin{proposition}\label{cor:optimal_is_reach}
\sbPROOR
\end{proposition}

\noindent The reduction $\textbf{MDP-CN} \leq_P \textbf{MDP-CN-qual}$
is described in procedure \FuncSty{Solve-CN}.
Its correctness follows from 
Proposition \ref{cor:optimal_is_reach}.
Once the set $A$ of vertices with $\Val^{\CNDONO{\D}{r}}=1$,
and a corresponding
$\CNDONO{\D}{r}$-optimal strategy, are both computed
(line~\ref{proc:solve-CN-qual-sub}, which calls the subroutine 
\FuncSty{Qual-CN} for solving \textbf{MDP-CN-qual}),
solving \textbf{MDP-CN} amounts to 
computing an MD strategy for maximizing the probability of
reaching a vertex in $A$, and computing the 
respective reachability probabilities.
This is done
on line~\ref{proc:solve-CN-max-reach} by calling procedure
\FuncSty{Max-Reach}.
It is well known that \FuncSty{Max-Reach} can be implemented in polynomial
time:  both an optimal strategy and the associated optimal (rational) 
probabilities
can be obtained by solving suitable linear programs 
(see, e.g., ~\cite{CY98} or ~\cite[Section 7.2.7]{Puterman94}).
Thus the running time of
\FuncSty{Solve-CN}, excluding the running time of \FuncSty{Qual-CN},
is polynomial.
Moreover, the optimal values are rational, so
Lemma~\ref{lem:equiv_OCDP_MDP} implies
that $\Val^{\CNCO{\A}}$ is also rational.

\subsection{Reduction to Qualitative $\MPONO{\D}{r}$}\label{subsec:qCN-to-qMP}
The reduction
$\textbf{MDP-CN-qual} \leq_P \textbf{MDP-MP-qual}$
is described in procedure \FuncSty{Qual-CN}.
Fixing some initial vertex $s$, let us denote by $\SMP{\D}{r}{\leq 0}$ the set
of all MD strategies $\sigma$ satisfying $\calP(\MPSNO{\sigma}{\D}{r}(s))=1$,
and by $\SCN{\D}$ the set
of all MD strategies $\sigma$ satisfying $\calP(\CNDSNO{\sigma}{\D}{r}(s))=1$.
It is not hard to see that
$
\SCN{\D}\subseteq\SMP{\D}{r}{\leq 0}
$.
If this was an equality, the reduction
would boil down to the identity map.
Unfortunately, these sets are not equal in general.
A trivial example is provided by a MDP with just one
vertex $s$ with reward $0$. More generally, the strategy
$\sigma$ may be trapped in a finite loop around $0$ (causing
$\calP(\MPSNO{\sigma}{\D}{r}(s))=1$) but never accumulate all negative values
(causing $\calP(\CNDSNO{\sigma}{\D}{r}(s))=0$). 
As a solution to this problem, we characterize in Lemma~\ref{lem:decreasing_MP_CN}
the strategies from $\SMP{\D}{r}{\leq 0}$ which are also in $\SCN{\D}$,
via the property
of being ``decreasing'':

\begin{procedure}[t]
\caption{Qual-CN($\D$,$r$)}\label{proc:qual-CN}
\dontprintsemicolon
\KwData{A MDP $\D$ with reward $r$.}
\KwResult{Compute the set $A\subseteq V$ of vertices with $\Val^{\CNDONO{\D}{r}}=1$,
and a MD strategy, $\sigma$, $\CNDONO{\D}{r}$-optimal starting at every $v\in A$.}

$\D'$ $\leftarrow$ \FuncSty{Decreasing}($\D$) \;
\nllabel{proc:qual-CN-decr}

$\left(A', \sigma'\right)$ $\leftarrow$ \FuncSty{Qual-MP}($\D'$,$r$) \;
\nllabel{proc:qual-CN-qual-sub}

$A \leftarrow \{v \in V \mid (v,1,0) \in A'\}$ \;

$\sigma$ $\leftarrow$ \FuncSty{CN-FD-to-MD}($\sigma'$) \;
\nllabel{proc:qual-CN-FD}

\Return{$\left(A, \sigma\right)$} \;
\end{procedure}

\begin{definition}
A MD strategy $\sigma$ in $\D$ is \emph{decreasing} if for every
state $u$ of $\D\langle\sigma\rangle$ reachable from $s$ there is
a finite path $w$ initiated in $u$ such that $r(w)=-1$.
\end{definition}

\begin{lemma}\label{lem:decreasing_MP_CN}
\sbLED
\end{lemma}

\begin{figure}[t]
\fbox{
\parbox{0.97\textwidth}{
$\D'=(V',\bbtran{},(V'_N,V'_P),\Prob')$, where
\begin{itemize}
\item $V'= \{(u,n,m), [u,n,m,v] \mid u \in V, u\btran{} v,0\leq n,m\leq |V|^2+1\} \cup \{\mathit{div}\}$
\item $V'_P = \{[u,n,m,v]\in V' \mid u\in V_P\}$, $V'_N=V'\smallsetminus V'_P$
\item transition relation $\bbtran{}$ is the \emph{least} set satisfying the following
  for every $u,v\in V$ such that $u\btran{} v$ and $0\leq m,n\leq |V|^2+1$:
  \begin{itemize}
  \item if $m=|V|^2+1$ and $n>0$, then $(u,n,m)\bbtran{}\mathit{div}$
  \item if $m\leq |V|^2+1$ and $n=0$, then $(u,n,m)\bbtran{}[u,1,0,v]$
  \item if $m<|V|^2+1$ and $n>0$, then $(u,n,m)\bbtran{}[u,n,m,v]$
  \item if $u\in V_P$, then $[u,n,m,v]\bbtran{}(v,n+r(u),m+1)$ 
	and
        $[u,n,m,v']\bbtran{} (v,1,0)$ 
	for all 
        $v'\in V\smallsetminus \{v\}$
        such that $[u,n,m,v']\in V'$
        
  \item if $u\in V_N$, then 
        $[u,n,m,v]\bbtran{}(v,n+r(u),m+1)$
  \item $\mathit{div}\bbtran{} \mathit{div}$
  \end{itemize} 
\end{itemize}
$\Prob'([u,n,m,v]\bbtran{} (v',n',m'))=\Prob(u\btran{} v')$ whenever
$[u,n,m,v]\in V'_P$ and $[u,n,m,v]\bbtran{} (v',n',m')$.
Finally, $r'((u,n,m))=0$, $r'([u,n,m,v])=r(u)$ and $r'(\mathit{div})=1$.
}}
\caption{Definition of the MDP $\D'$.}\label{fig:MDP_Dprime}
\end{figure}

A key part of the reduction is the construction
of an MDP, $\D'$, described in Figure~\ref{fig:MDP_Dprime},
which simulates the MDP $\D$, but satisfies that
$\SMP{\D'}{r}{\leq 0}=\SCN{\D'}$ for every initial vertex $s$.
The idea is to augment the vertices of $\D$ with additional
information, keeping track of whether the run under some
$\sigma \in \SMP{\D'}{r}{\leq 0}$
``oscillates'' with accumulated rewards in a bounded neighborhood
of $0$, or ``makes progress'' towards $-\infty$.
The last obstacle in the reduction is that MD strategies
for $\D'$ do not directly yield MD strategies for $\D$.
Rather a $\CNDONO{\D'}{r}$-optimal MD strategy, $\tau'$, for $\D'$
induces a deterministic $\CNDONO{\D}{r}$-optimal strategy, $\tau$, which
uses a finite automaton to evaluate the history of play.
Fortunately, given such a strategy $\tau$ it is possible to
transform it to a $\CNDONO{\D}{r}$-optimal MD strategy for $\D$
by carefully eliminating the memory it uses.
This is done on line~\ref{proc:qual-CN-FD}.
We postpone the proof of these claims
to the  Appendix, and just note that the construction
of $\D'$ on line~\ref{proc:qual-CN-decr}, procedure \FuncSty{Decreasing}
can clearly be done in polynomial time.  Thus, the overall time
complexity of the reduction is polynomial.

\vspace*{-0.08in}

\subsection{Solving Qualitative $\MPONO{\D}{r}$}\label{subsec:solve-MP}

For a fixed vertex $s \in V$, for
every MD strategy $\sigma$ and reward function
$r$, we define a random variable $\rwV{\sigma}{r}{s}$ such that
for every run $w\in \run_{\D\langle\sigma\rangle}(s)$:

\vspace*{-0.18in}

$$
\rwV{\sigma}{r}{s}(w) = \begin{cases}
       \lim_{n\to \infty} \frac{r({w}\downarrow{n})}{n} & \text{ if the limit exists;} \\
       \bot               & \text{ otherwise.}
       \end{cases}
$$
It follows from, e.g., \cite[Theorem~1.10.2]{Norris98} that
since $\sigma$ is MD
the value of $\rwV{\sigma}{r}{s}$ is almost surely defined.
Solving the $\MPONO{\D}{r}$ objective amounts to
finding a MD strategy $\sigma$ such that
$\calP(\rwV{\sigma}{r}{s} \leq 0)$ is maximal among
all MD strategies.
We use the procedure \FuncSty{get-MD-min} to find
for every vertex $s\in V$ and a reward
function $r$ a MD strategy $\varrho$
such that
$E\rwV{\varrho}{r}{s}=\min_{\sigma \in MD}E\rwV{\sigma}{r}{s}$.
This can be done in polynomial time via linear programming: see, e.g., 
~\cite[Algorithm~2.9.1]{FilarVrieze} or \cite[Section 9.3]{Puterman94}.

\begin{procedure}[t]
\caption{Qual-MP($\D$,$r$)}\label{proc:mp-qual}
\dontprintsemicolon
\KwData{A MDP $\D$ with reward $r$.}
\KwResult{Compute the set $A\subseteq V$ of vertices with $\Val^{\MPONO{\D}{r}}=1$
and a MD strategy $\sigma$ $\MPONO{\D}{r}$-optimal starting in every $v\in A$.}

$V_?\leftarrow V$, $A\leftarrow \emptyset$, $T\leftarrow \emptyset$, $\hat{r}\leftarrow r$ \;
\nllabel{proc:mp-qual-init}

\While{$V_? \neq \emptyset$}%
{

$s$ $\leftarrow$ \FuncSty{Extract}($V_?$) \;
\nllabel{proc:mp-qual-s}

\If{$\exists \varrho: E\rwV{\varrho}{\hat{r}}{s}\leq 0$}%
{

$\varrho$ $\leftarrow$ \FuncSty{get-MD-min}($\D$,$r$,$s$) \;
\nllabel{proc:mp-qual-varrho} 

$C$ $\leftarrow$ a BSCC $C$ of $\D\langle\varrho\rangle$
such that $C\cap A=\emptyset$
and $\calP(\rwV{\varrho}{\hat{r}}{s}\leq 0 \mid {Reach}^{\varrho}_{C})=1$ \;
\nllabel{proc:mp-qual-C}

$\left(\tau,\left(reach_v\right)_{v\in V}\right)$ $\leftarrow$ \FuncSty{Max-Reach}($\D$,$C \cup A$) \;
\nllabel{proc:mp-qual-tau}

$A' \leftarrow \{u \in V \mid reach_u=1\}$ \;
\nllabel{proc:mp-qual-Ap}

\lFor{every $u \in V_N, v \in V$}%
{
\lIf{$\left(u\in C \land v=\varrho(u)\right) \lor \left(u\in A'\smallsetminus (C\cup A) \land v=\tau(u)\right)$}%
{$T\leftarrow T \cup \{(u,v)\}$
}%
}%
\;
\nllabel{proc:mp-qual-T}

$A\leftarrow A' \cup A$  \;
\nllabel{proc:mp-qual-A-next}

\lFor{every $u \in V$}%
{\lIf{$u\in A$}{$\hat{r}(u)\leftarrow 0$}
}
\;
\nllabel{proc:mp-qual-r-next}

\lIf{$s \notin A$}{$V_? \leftarrow V_?\cup\{s\}$} \;
\nllabel{proc:mp-qual-s-back}

}%
}%

$\sigma$ $\leftarrow$ \FuncSty{MD-from-edges}($T$) \;
\nllabel{proc:mp-qual-sigma}

\Return{($A$, $\sigma$)} \;
\nllabel{proc:mp-qual-fin}

\end{procedure}

The core idea of procedure \FuncSty{Qual-MP} for solving \textbf{MDP-MP-qual}
is this:
Whenever $E\rwV{\tau}{r}{s}\leq 0$ then there is a bottom strongly connected
component (BSCC), $C$, of the transition graph of $\D\langle\tau\rangle$,
such that almost all runs $w$ reaching $C$ satisfy $\rwV{\tau}{r}{s}(w)\leq 0$.
Since $\Val^{\MPONO{\D}{r}}(s)=1$ implies the existence of some $\tau\in\SMP{\D}{r}{\leq 0}$
such that $E\rwV{\tau}{r}{s}\leq 0$,
\FuncSty{Qual-MP} solves \textbf{MDP-MP-qual}
by successively cutting off the BSCCs just mentioned,
while maintaining the invariant
$\exists \tau:E\rwV{\tau}{r}{s}\leq 0$.
Details and proofs are in the Appendix.

\FuncSty{Extract}($S$) removes an arbitrary element
of a nonempty set $S$ and returns it,
and \FuncSty{MD-from-edges}($T$)
returns an arbitrary MD strategy $\sigma$ satisfying $(u,v)\in T \land u\in V_N \Rightarrow \sigma(u)=v$.
Both these procedures can clearly be implemented in polynomial time.
Thus by the earlier discussion about the complexity of \FuncSty{Max-Reach},
in Section~\ref{subsec:CN-to-qual}, we conclude that
\FuncSty{Qual-MP} runs in polynomial time.

%% file: bound.tex
\section{OC-MDPs with Boundary}
\label{sec-boundary}

\newcommand{\FOREACH}{\textbf{for each }}
\newcommand{\IF}{\textbf{if }}
\newcommand{\THEN}{\textbf{then }}
\newcommand{\ELSE}{\textbf{else }}
\newcommand{\REPEAT}{\textbf{repeat }}
\newcommand{\UNTIL}{\textbf{until }}
\newcommand{\DO}{\textbf{do }}
\newcommand{\DONE}{\textbf{done }}
\newcommand{\CONTINUE}{\textbf{continue }}
\newcommand{\OUTPUT}{\textbf{output }}
\newcommand{\cc}{\mathtt{check\_color}}
\newcommand{\cp}{\mathtt{check\_path}}
\newcommand{\cv}{\mathtt{check\_value}}
\newcommand{\col}{\mathit{col}}
\newcommand{\fin}{\mathit{fin}}
\newcommand{\pconf}{\mathit{PathConf}}
\newcommand{\Black}{\mathit{Black}}
\newcommand{\White}{\mathit{White}}

Fix an OC-MDP,
$\A = (Q,\delta^{=0},\delta^{>0},(Q_N,Q_P),P^{=0},P^{>0})$,
and its associated MDP, $\Dz_\A$.

\begin{definition}[termination objectives]
\label{def-term-obj}
  The (non-selective) \emph{termination objective}, denoted $\NT$, 
  consists of all runs $w$ of $\Dz_\A$ that eventually 
  hit a configuration with counter
  value zero. 
  Similarly, for a set $F \subseteq Q$ of 
  \emph{final} states we define the associated 
  \emph{selective termination objective}, denoted $\ST_F$
  (or just $\ST$ if $F$ is understood), consisting of all
  runs of $\Dz_\A$ that hit a configuration of the form $q(0)$ where $q \in F$.
\end{definition}

\noindent
Termination objectives are more complicated than the $\CN$ objectives
considered in Section~\ref{sec-noboundary}, and even 
\emph{qualitative} problems for them require new insights.
We define $\ValOne^{\NT}$ and $\ValOne^{\ST}$ be the sets of all $p(i) \in
Q {\times} \Nset_0$ such that $\Val^{\NT}(p(i)) =1$ and
$\Val^{\ST}(p(i)) =1$, respectively. We also define their subsets
$\OptValOne^{\NT}$ and $\OptValOne^{\ST}$ consisting of all
$p(i) \in \ValOne^{\NT}$ and all $p(i) \in \ValOne^{\ST}$, respectively, 
such that 
there is an optimal strategy achieving value
1 starting at $p(i)$.
Are the inclusions
$\OptValOne^{\NT} \subseteq \ValOne^{\NT}$ and
$\OptValOne^{\ST} \subseteq \ValOne^{\ST}$
proper? 
It turns out that the two objectives differ in this respect.
We begin by stating our results about qualitative
$\NT$ objectives. 

\begin{theorem}
\label{thm-NT}
  $\ValOne^{\NT} = \OptValOne^{\NT}$. Moreover, 
  given a OC-MDP, $\A$, and a configuration $q(i)$ of $\A$,
  we can decide in polynomial time whether $q(i) \in \ValOne^{NT}$. 
  Furthermore, there is a $\SMD$ 
  strategy, $\sigma$, constructible in polynomial time, which is 
  optimal starting at every configuration in $\ValOne^{\NT} = \OptValOne^{\NT}$.
\end{theorem}

Next we turn to $\ST$ objectives. 
First, the inclusion $\OptValOne^{\ST} \subseteq \ValOne^{\ST}$ is
proper: there may be no optimal strategy for $\ST$ even when
the value is 1. See Appendix \ref{app-boundary} for
an example that establishes this.
We provide an exponential time algorithm to
decide whether a given configuration $q(i)$ is in  
$\OptValOne^{\ST}$, and we show there is a 
``counter-regular''
strategy $\sigma$ constructible in exponential time that is optimal
starting at all configurations in $\OptValOne^{\ST}$. 
We first introduce the notion of \emph{coloring}. 

\begin{definition}[coloring]
\label{def-coloring}
A \emph{coloring} is a map 
\mbox{$C : Q\times\Nset_0 \rightarrow \{b,w,g,r\}$}, where 
$b$, $w$, $g$, and $r$ are the four different ``colors'' 
(black, white, gray, and red).
For every $i \in \Nset_0$, we define the \emph{$i$-th column} of $C$ as 
a map $C_i : Q \rightarrow \{b,w,g,r\}$, where $C_i(q) = C(q(i))$. 
\end{definition}

\noindent
A coloring can be depicted as an infinite matrix of points (each being 
black, white, gray, or red) with rows indexed by control states
and columns indexed by counter values. We are mainly interested
in the coloring, $R$, which represents the set $\OptValOne^{\ST}$ in the
sense that for every $p(i) \in Q \times \Nset_0$, the value of $R(p(i))$ 
is either $b$ or $w$, depending on whether $p(i) \in \OptValOne^{\ST}$ 
or not. First, we show $R$ is 
``ultimately periodic'':

\begin{lemma}
\label{lem-period}
Let  $N = 2^{|Q|}$.
  There is an $\ell$, $1 \leq \ell \leq N$, such that for  
  $j \geq N$, we have $R_j = R_{j+\ell}$. 
\end{lemma}

\noindent
Thus the coloring $R$ consists of an
``initial rectangle'' of width \mbox{$N + 1$} followed by infinitely
many copies of the ``periodic rectangle'' of width $\ell$ (see
Fig.~\ref{fig-R-coloring} in appendix \ref{app-boundary}).  
Note that $R_N = R_{N+\ell}$.
We show how to compute the initial and periodic 
rectangles of~$R$ by, intuitively, trying out all
(exponentially many) candidates for the width $\ell$ and the 
columns $R_N = R_{N+\ell}$. For each such pair of candidates, the algorithm
tries to determine the color of the remaining points in the initial
and periodic rectangles, until it either finds an inconsistency
with the current candidates, 
or produces a coloring which is
not necessarily the same as $R$, but where all black points are certified 
by an optimal strategy. Since the algorithm eventually tries also
the ``real'' $\ell$ and $R_N = R_{N+\ell}$, all black points of $R$
are discovered. We note that the polynomial-time
algorithm for $\CN$ objectives 
is used as a ``black-box'' here and applied
to various OC-MDPs constructed from $\A$ and the current coloring
maintained by the algorithm (see Fig.~\ref{fig-alg-R}). The
many subtleties are discussed
in Appendix~\ref{app-boundary}.

\begin{theorem}
\label{thm-optvalone}
  An automaton recognizing $\OptValOne^{\ST}$, and a counter-regular strategy, $\sigma$,
  optimal starting at very
configuration in $\OptValOne^{\ST}$,
 are both computable in exponential time.
\end{theorem}

\noindent
Thus, membership in
$\OptValOne^{\ST}$ is solvable in exponential time.
We do not have an analogous result for $\ValOne^{\ST}$ and
leave this as an open problem
(the
example in appendix \ref{app-boundary} gives a taste of the
difficulties). 

A straightforward reduction from the
emptiness problem for alternating finite automata over a
one-letter alphabet, which is $\PSPACE$-hard,
see e.g.~\cite{JS07},
shows that membership in $\OptValOne^\ST$ is $\PSPACE$-hard.

Further, we show
that membership in
$\ValOne^{\ST}$
is hard for the Boolean Hierarchy (\textbf{BH}) over $\NP$, and thus
neither in  $\NP$ nor $\coNP$ assuming standard complexity assumptions.
The proof technique, based on a number-theoretic encoding, originated in 
\cite{Kuc03} and was used in
\cite{JKMS04,Serre06}.

\begin{theorem}
\label{thm-term-hard}
  Membership in $\ValOne^{\ST}$ is \textbf{BH}-hard.
  Membership in $\OptValOne^{\ST}$ is $\PSPACE$-hard.
\end{theorem}

As noted in the introduction, for the
very special subclass of {\em solvency games} 
\cite{BKSV08}, all {\em qualitative} problems are decidable in polynomial time
(see Appendix \ref{app-boundary} for formal definitions and proofs):

\begin{proposition}
\label{prop-solvency}
Given a solvency game, it is decidable in polynomial time
whether the gambler has a strategy to go bankrupt with  
probability: \ \ $> 0$, \ $=1$, \ $=0$, or \ $<1$.
\end{proposition}

The cases other than $< 1$ are either trivial or follow 
easily from what we have established 
for OC-MDPs. For the case $< 1$, we make use of a lovely theorem 
on inhomogeneous (controlled) random
walks \cite{DKL91}.

\paragraph{Acknowledgement.}
The authors thank Petr Jan\v{c}ar, Richard Mayr, and Olivier Serre
for pointing out the $\PSPACE$-hardness of the membership problem
for $\OptValOne^{\ST}$.

%% file: app-nobound.tex
\section{Proofs of \protect{Section~\ref{sec-noboundary}}}
\label{app-noboundary}

\subsection{Proof of Lemma~\ref{lem:equiv_OCDP_MDP}}
\begin{reftheorem}{Lemma}{\ref{lem:equiv_OCDP_MDP}}
\sbLEOM
\end{reftheorem}

\begin{proof}
Consider a MDP $\D = 
(Q\cup \delta^{>0},\btran{},(Q_N\cup \delta^{>0},Q_P),\Prob)$ where 
\[
\btran{}:=\{(p,(p,d,q))\mid (p,d,q)\in \delta^{>0}\}\cup \{((p,d,q),q)\mid
(p,d,q)\in \delta^{>0}\}
\] 
and $\Prob(p,(p,d,q))=P^{>0}(p,d,q)$ for every $p\in Q_P$.
Consider a reward function $r:(Q\cup \delta^{>0})\to \{-1,0,1\}$ such that
$r(p)=0$ for $p\in Q$, and $r((p,d,q))=d$ for $(p,d,q)\in \delta^{>0}$.

Consider $\Dn_\A=(Q \times \Zset,{\ctran{}},(Q_N \times \Zset,Q_P \times \Zset),\Prob)$.
Let $\Theta$ be a mapping of paths in $\D$ to paths in $\Dn_\A$ defined as follows:
Given a finite path $\omega=p_1(p_1,d_1,p_2)p_2(p_2,d_2,p_3)\cdots 
(p_{n-1},d_{n-1},p_n) p_n$
in $\D$, we define $\Theta(\omega)$ to be the path
$p_1(i) p_2(i+d_1)\cdots p_n(i+\sum_{j=1}^{n-1} d_j)$.
Observe that the mapping is one-to-one and onto.

Let $\bar{\sigma}$ be a HR strategy in $\Dn_\A$. We define a strategy $\sigma$ in $\D$
as follows: For every path $\omega=p_1(p_1,d_1,p_2)p_2(p_2,d_2,p_3)\cdots 
(p_{n-1},d_{n-1},p_n) p_n$ in $\D$ we have that
$\sigma(\omega)$ assigns $x$ to a transition $(p_n,(p_n,d,q))$ iff
$\bar{\sigma}(\Theta(\omega))$ assigns $x$ to 
$(p_n(i+\sum_{j=1}^{n-1} d_j),q(i+\sum_{j=1}^{n-1} d_j+d))$.
Let us extend $\Theta$ to runs $w\in \run_{\D(\sigma)}(p)$ by
$\Theta(w)(i)=\Theta(w(2i))$.
Then $\Theta: \run_{\D(\sigma)}(p) \to \run_{\Dn_\A(\bar{\sigma})}(p(i))$
is a bijection
and induces an isomorphism of the corresponding probability spaces.%
\footnote{I.e. for any $A \subseteq \run_{\D(\sigma)}(p)$ we have
that $A$ is measurable iff $\Theta(A)$ is measurable
and $\calP(A)=\calP(\Theta(A))$.}
Also,
$\Theta(\CNDSNO{\sigma}{\D}{r}(p))=\CNCS{\bar{\sigma}}{\A}(p(i))$.
Thus $\calP(\CNDSNO{\sigma}{\D}{r}(p))=
\calP(\CNCS{\bar{\sigma}}{\A}(p(i)))$, and hence
$\Val^{\CNDONO{\D}{r}}(p)\geq \Val^{\CNCO{\A}}(p(i))$ because
$\bar{\sigma}$ was arbitrary.

Let $\sigma$ be a HR strategy in $\D$. We define a strategy $\bar{\sigma}$ in
$\Dn_\A$ as follows: For every path 
$\omega'=p_1(i) p_2(i+d_1)\cdots p_n(i+\sum_{j=1}^{n-1} d_j)$ in $\Dn_\A$
we have that $\bar{\sigma}(\omega')$ 
assigns $x$ to 
$(p_n(i+\sum_{j=1}^{n-1} d_j),q(i+\sum_{j=1}^{n-1} d_j+d))$
iff $\sigma(\Theta^{-1}(\omega'))$ assigns $x$ to $(p_n,(p_n,d,q))$.
Similarly as above,
$\calP(\CNDSNO{\sigma}{\D}{r}(p))=
\calP(\Theta(\CNDSNO{\sigma}{\D}{r}(p)))=
\calP(\CNCS{\bar{\sigma}}{\A}(p(i)))$.
It follows that $\Val^{\CNDONO{\D}{r}}(p)\leq \Val^{\CNCO{\A}}(p(i))$
because $\sigma$ was arbitrary. This finishes the proof of 1.

For 2., note that if $\sigma$ is a MD strategy, then the strategy
$\bar{\sigma}$ defined in the previous paragraph coincides with
the strategy $\sigma'$ from the statement of the lemma on
paths of $\Dn_\A$. However, then
$\calP(\CNDSNO{\sigma}{\D}{r}(p))=
\calP(\CNCS{\bar{\sigma}}{\A}(p(i)))=
\calP(\CNCS{\sigma'}{\A}(p(i)))$.
\end{proof}

\subsection{Proof of existence of CN-optimal MD strategies}%
\label{proof:prop:CN_optimal_RewMDP}
We prove that the existence of a $\CNDONO{\D}{r}$-optimal 
MD strategy
for finite-state MDPs with rewards
follows from
\cite[Theorem~1]{Gimbert-STACS07}.
To do so
we need to introduce the following notions from \cite{Gimbert-STACS07}.
Note that the notions are simplified to achieve an easier formulation
but all the arguments can be easily modified to use the original notions.

Let $O\subseteq \run_{\D}$ be a measurable objective.
We say that $O$ is \emph{positional} if 
there is some MD strategy
$\bar\sigma$ such that
every $v\in V$ satisfies $\calP(O^{\bar{\sigma}}(v))=
\sup_{\sigma\in {HR}} \calP(O^{\sigma}(v))$.
Moreover $O$ is \emph{prefix independent} if for every
run $w\in \run_{\D}$ and every finite path $w'$ such that
$w'w$ is a run we have that 
$w\in O$ iff $w'w\in O$. %
Finally, $O$ is \emph{submixing} if for
every infinite sequence of finite paths $u_0, v_0, u_1, v_1, \ldots$
such that $u_0 v_0 u_1 v_1 \cdots$,
$u_0 u_1 \cdots$ and
$v_0 v_1 \cdots$ are runs the following is true:
If $u_0 v_0 u_1 v_1 \cdots\in O$, then
$u_0 u_1 \cdots\in O$, or $v_0 v_1 \cdots\in O$.
Theorem~1 of \cite{Gimbert-STACS07} implies that every
prefix independent submixing objective is positional\footnote{Note
that the results of \cite{Gimbert-STACS07} are more general and consider
measurable pay-off functions on runs instead of sets of runs.
However, if $O$ is prefix-independent and submixing according
to the definition given here, then clearly
the characteristic function of $O$ is a
prefix independent and submixing pay-off function, as defined in
\cite{Gimbert-STACS07}, and hence the results of
\cite{Gimbert-STACS07} apply.}.

$\CNDONO{\D}{r}$ is clearly prefix independent.
We now prove that it is also submixing.
Let $w=u_0 v_0 u_1 v_1 \cdots$ be a run. 
For $n\in\Nset$ we denote $u{\lfloor} n$ the subword of $w{\downarrow}n$
obtained by leaving out all $v_i$-parts.
Similarly we denote $v{\lfloor}n$ the subword of $w{\downarrow}n$
obtained by leaving out all $u_i$-parts.
Note that
$r(w{\downarrow}n) = r(u{\lfloor}n)+r(v{\lfloor}n)$. However, then
clearly either $\liminf_{n\to\infty}(u{\lfloor}n)=-\infty$, or
$\liminf_{n\to\infty}(v{\lfloor}n)=-\infty$. It follows that
either $u_0 u_1\cdots \in \CNDONO{\D}{r}$, or $v_0 v_1\cdots\in \CNDONO{\D}{r}$, i.e.,
$\CNDONO{\D}{r}$ is submixing.
We therefore have:

\begin{lemma}[cf. \cite{Gimbert-STACS07}]
\label{lem:CN-opt-MD-exists}For finite-state MDPs with rewards,
there always exists a $\CNDONO{\D}{r}$-optimal
MD strategy.
\end{lemma}

\subsection{Auxiliary lemma concerning CN objectives and MD strategies}
\label{subsec:auxiliary_CN_MD}
\begin{lemma}\label{lem:mp_nonpos}
Let $\sigma$ be a MD strategy in $\D$ and
let $C$ be a bottom strongly connected component (BSCC) of $\D\langle\sigma\rangle$. Given $u\in C$, we define
$R^{\sigma}_u:\run_{\D\langle\sigma\rangle}(u)\to \Rset$
to be a random variable giving the reward accumulated before the run returns to $u$, i.e.,
\[R^{\sigma}_u(w)=\begin{cases}
      r(w{\downarrow}n) & \text{ if }
                         n=\min\{j\geq 1\mid w(j)=u\}<\infty \\
      \infty & \text{ otherwise}
      \end{cases}\]
Then there is $x_C\in \{0,1\}$ such that for all $u\in C$ we have
$\calP(\CNDSNO{\sigma}{\D}{r}(u))=x_C$.
Moreover, $x_C=1$ iff for \emph{some} $u\in C$ we have
$\calP(R^{\sigma}_u<0)>0$ and $ER^{\sigma}_u\leq 0$
(here $ER^{\sigma}_u$ is the expected value of $R^{\sigma}_u$).
\end{lemma}

\begin{proof} 
Let us fix $u\in C$.
From~\cite[Theorem~1.10.2]{Norris98} we have that $\calP({Reach}^{\sigma}_{\{u\}}(v))=1$ for all $v\in C$.
Thus we have
$\calP(\CNDSNO{\sigma}{\D}{r}(u))=\calP(\CNDSNO{\sigma}{\D}{r}(v))$
because $\CN$ is prefix independent, moreover
$\calP(R^{\sigma}_u=\infty)=0$.
Hence, it suffices to show
that $\calP(\CNDSNO{\sigma}{\D}{r}(u))\in \{0,1\}$, and
that $\calP(\CNDSNO{\sigma}{\D}{r}(u))=1$ iff 
$\calP(R^{\sigma}_u<0)>0$ and $ER^{\sigma}_u\leq 0$.

We define sequences of random variables $I_1,I_2,I_3\ldots$
and $X_1,X_2,\ldots$ as follows: given a run $w\in\run_{\D\langle\sigma\rangle}(u)$, 
we define $I_1(w)=0$, and for all $n\geq 2$ we define 
$I_n(w)$ to be the least $m>I_{n-1}(w)$ such that $w(m)=u$.
We define $X_n(w)=r(w{\downarrow}I_{n+1}(w))-r(w{\downarrow}{I_{n}(w)})$ 
the reward accumulated between 
the $n$-th visit to $u$ (inclusive) and $n+1$-th visit to $u$ (non-inclusive). 
Observe that $X_1=R^{\sigma}_u$ and that
the variables $X_1,X_2,\ldots$
are identically distributed
and independent.
Therefore, the sequence $X_1,X_2,\ldots$ determines a random walk
$S_0,S_1,S_2,\ldots$ on $\Zset$ where $S_n=\sum_{i=1}^n X_i$.

Suppose that $\calP(R^{\sigma}_u<0)>0$ and $ER^{\sigma}_u\leq 0$.
There are two cases depending on 
whether $\calP(R^{\sigma}_u>0)=0$, or not.
First, assume that $\calP(R^{\sigma}_u>0)=0$ and thus also $EX_1=EX_j<0$ for all $j$.
Then almost all $w\in \run_{\D\langle\sigma\rangle}(u)$ satisfy the following:
$X_i(w)\leq 0$ for every $i\geq 0$, and $X_j(w)<0$ for infinitely many $j\geq 0$,
as follows from the strong law of large numbers,
see~e.g.~\cite[Theorem~22.1]{Billing95}, and
the fact that $EX_j<0$. However,
then $\calP(\CNDSNO{\sigma}{\D}{r})=1$.
Now assume that $\calP(R^{\sigma}_u>0)>0$.
We may apply, e.g., \cite[Theorem 8.3.4]{Chung01} and conclude that
almost all $w\in \run_{\D\langle\sigma\rangle}(u)$ satisfy 
$\liminf_{n\to \infty} S_n(w)=-\infty$, 
which implies that 
$\calP(\CNDSNO{\sigma}{\D}{r})=1$.

Now suppose that either $\calP(R^{\sigma}_u<0)>0$, or $ER^{\sigma}_u\leq 0$
is not satisfied. If $\calP(R^{\sigma}_u<0)=0$, then clearly for all
$w\in \run_{\D\langle\sigma\rangle}(u)$ and
for every $n\geq 0$ we have $r(w{\downarrow} n)\geq -|V|$, 
which implies that $\calP(\CNDSNO{\sigma}{\D}{r})=0$.
If $\calP(R^{\sigma}_u<0)>0$ but $ER^{\sigma}_u>0$, then
using, e.g., \cite[Theorem 8.3.4]{Chung01},
almost all $w\in \run_{\D\langle\sigma\rangle}(u)$ satisfy 
$\lim_{n\to \infty} S_n(w)=\infty$, 
which implies that $\calP(\CNDSNO{\sigma}{\D}{r})=0$.
\end{proof}

\subsection{Proof of Proposition~\ref{cor:optimal_is_reach}}
\begin{reftheorem}{Proposition}{\ref{cor:optimal_is_reach}}
\sbPROOR
\end{reftheorem}

\begin{proof}
The fact that $\max_{\tau\in {MD}}\calP({Reach}^{\tau}_{A}(u))=
\sup_{\tau\in {HR}}\calP({Reach}^{\tau}_{A}(u))$ follows from~\cite[Section 7.2.7]{Puterman94},
see also~\cite{CY98}.
Clearly $\max_{\tau\in {MD}}\calP({Reach}^{\tau}_{A}(u))\leq 
\Val^{\CNDONO{\D}{r}}(u)$. For the opposite direction,
let us pick a $\CNDONO{\D}{r}$-optimal MD strategy $\sigma$.
Consider the Markov chain
$\D\langle\sigma\rangle$ with states $V$. By Lemma~\ref{lem:mp_nonpos} (see Section~\ref{subsec:auxiliary_CN_MD}), for every BSCC 
$C$ of
$\D\langle\sigma\rangle$ there is a number $x_C\in\{0,1\}$ such that
$x_C = \calP(\CNDSNO{\sigma}{\D}{r}(v)) = \Val^{\CNDONO{\D}{r}}(v)$ 
for all $v\in C$.
Let us denote by $\mathcal{C}$ the union of all BSCCs $C$ such that $x_C=1$.
Let $\pi$ be a MD strategy such that
$\calP({Reach}^{\pi}_{\C}(u))= 
\max_{\tau\in {MD}}\calP({Reach}^{\tau}_{\C}(u))$. Then
$\calP(\CNDSNO{\sigma}{\D}{r}(u))\leq\calP({Reach}^{\pi}_{\mathcal{C}}(u))$ because
almost all runs of $\D\langle\sigma\rangle$ eventually reach a BSCC.
However, $\mathcal{C}\subseteq A$, and thus
\[
\Val^{\CNDONO{\D}{r}}(u)=
\calP(\CNDSNO{\sigma}{\D}{r}(u))\leq\calP({Reach}^{\pi}_{\mathcal{C}}(u))
\leq \calP({Reach}^{\pi}_{A}(u))\leq
\max_{\tau\in {MD}}\calP({Reach}^{\tau}_{A}(u))
\]
\end{proof}

\subsection{Proof of Lemma~\ref{lem:decreasing_MP_CN}}
We fix an arbitrary initial state $s$ and consider
the sets of strategies $\SMP{\D}{r}{\leq 0}$ and $\SCN{\D}$ defined
with respect to $s$, see Section~\ref{subsec:qCN-to-qMP}.
Recall that
a MD strategy $\sigma$ in $\D$ is \emph{decreasing} if for every
state $u$ of $\D\langle\sigma\rangle$ reachable from $s$ there is
a finite path $w$ initiated in $u$ such that $r(w)=-1$.
We restate and prove Lemma~\ref{lem:decreasing_MP_CN} here.
\begin{reftheorem}{Lemma}{\ref{lem:decreasing_MP_CN}}
\sbLED
\end{reftheorem}

\begin{proof}
Let $\sigma$ be a MD strategy.
Denote $\C$ the union of all BSCCs
of $\D\langle\sigma\rangle$ reachable from $s$.
From~\cite[Theorem~1.10.2]{Norris98} we have that $\calP({Reach}^{\sigma}_\C(s))=1$
Let $u\in \C$.
Similarly as in the proof of Lemma~\ref{lem:mp_nonpos} (see Section~\ref{subsec:auxiliary_CN_MD}), 
we define sequences of random variables $I_1,I_2,I_3\ldots$
and $X_1,X_2,\ldots$ as follows: given a run $w\in\run_{\D\langle\sigma\rangle}(u)$, 
we define $I_1(w)=0$, and for all $n\geq 2$ we define 
$I_n(w)$ to be the least $m>I_{n-1}(w)$ such that $w(m)=u$.
We define $X_n(w)=r(w{\downarrow}I_{n+1}(w))-r(w{\downarrow}{I_{n}(w)})$ 
the reward accumulated between 
the $n$-th visit to $u$ (inclusive) and $n+1$-th visit to $u$ (non-inclusive). 
Observe that $X_1=R^{\sigma}_u$. We define $D_n=I_{n+1}(w)-I_n(w)$.
Observe that both $X_1,X_2,\ldots$ and $D_1,D_2,\ldots$
are sequences of identically distributed
and independent random variables.
Also $EX_1$ is finite, $0<ED_1<\infty$, and $X_1=R^{\sigma}_u$ where
$R^{\sigma}_u$ is the variable defined in Lemma~\ref{lem:mp_nonpos}.
By the strong law of large numbers, 
for almost all $w\in \run_{\D\langle\sigma\rangle}(u)$ 
\[
ER^{\sigma}_u=EX_1=\lim_{n\to \infty} \frac{\sum_{i=1}^{n} X_i(w)}{n}=
\lim_{n\to \infty} \frac{\sum_{i=1}^{n} X_i(w)}{\sum_{i=1}^{n} D_i(w)}
\frac{\sum_{i=1}^{n} D_i(w)}{n}=
\lim_{n\to \infty} \frac{\sum_{i=1}^n r(w(i))}{n} ED_1
\]
Assume that $\sigma\in \SCN{\D}$. Let $u\in \C$. We have
$\calP(\CNDSNO{\sigma}{\D}{r}(u))=1$ because $\CN$ is prefix independent
and $u$ is reachable from $s$.
Then, by Lemma~\ref{lem:mp_nonpos}, $ER^{\sigma}_u\leq 0$, and hence 
$\calP(\MPSNO{\sigma}{\D}{r}(u))=1$ by the above equation.
It follows that $\sigma\in \SMP{\D}{r}{\leq 0}$ 
because $u$ was an arbitrary state of $\C$, 
almost all runs initiated in $s$ reach $\C$,
and $\MP$ is prefix independent.

Assume that $\sigma\in \SMP{\D}{r}{\leq 0}$ and that $\sigma$ is decreasing. Let $u\in \C$. We have
$\calP(\MPSNO{\sigma}{\D}{r}(u))=1$ because $\MP$ is prefix independent
and $u$ is reachable from $s$.
Then, by the above equation, $ER^{\sigma}_u\leq 0$. Also,
$\calP(R^{\sigma}_u<0)>0$ because $\sigma$ is decreasing.
Hence, by Lemma~\ref{lem:mp_nonpos},
$\calP(\CNDSNO{\sigma}{\D}{r}(u))=1$.
It follows that $\sigma\in \SCN{\D}$ 
because $u$ was an arbitrary state of $\C$, 
almost all runs initiated in $s$ reach $\C$,
and $\CN$ is prefix independent.
\end{proof}

\subsection{Properties of $\D'$ and the correctness of \FuncSty{Qual-CN}}
Recall the MDP $\D'$ from Figure~\ref{fig:MDP_Dprime}. In this section
we prove some of its properties and prove that the procedure \FuncSty{Qual-CN}
from Section~\ref{subsec:qCN-to-qMP} is correct.
Also recall that whenever we use the sets $\SCN{\D}$ and $\SMP{\D}{r}{\leq 0}$
an initial vertex $s$ has to be specified, see the definition
of the sets in Section~\ref{subsec:qCN-to-qMP}.

\begin{lemma}\label{lem:short_path}
Let an initial vertex $s \in V$ be fixed and let $\sigma\in \SCN{\D}$. For every 
state $u$ of $\D\langle\sigma\rangle$ there is
a finite path $w$ of length at most $|V|^2+1$ initiated in $u$ such that $r(w)=-1$.
\end{lemma}

\begin{proof}
Let $w$ be the shortest path initiated in $u$ such that $r(w)=-1$.
Observe that if there are $i<j$ such that $w(i)=w(j)$ and
$r(w{\downarrow}i)\leq r(w{\downarrow}j)$, then
the path is not the shortest one (consider the path
$w(0)\cdots w(i) w(j+1)\cdots$). However, then every vertex can occur at most $|V|$
times in $w{\downarrow}(\len(w)-1)$. This gives $|V|^2+1$ 
upper bound on the length of $w$.
\end{proof}

Before we proceed to formal treatment, we briefly explain the intuition
behind the construction of $\D'$.
We start with explaining what information is kept in the vertices of $\D'$.
In what follows, vertices of the form $(u,1,0)$ for some $u\in V$
are called \emph{checkpoints}.
\begin{itemize}
\item First coordinate: the current vertex of $\D$;
\item second coordinate: the number by which the counter has to be decreased
to make the sum of rewards gained since the last checkpoint negative;
\item third coordinate: the number of steps since the last checkpoint;
\item fourth coordinate, if present: the next vertex of $\D$ through which the ``short path''
from the last checkpoint, see Lemma~\ref{lem:short_path}, should continue.
\end{itemize}
When the run starts, the first counter in the current vertex is
$1$ indicating that we wait for the sum of rewards becoming $-1$,
and the counter of steps is set to $0$. As the play proceeds,
the counters are updated accordingly. Whenever the first counter reaches
value zero, the play reaches a checkpoint and the counters
are reset to $1$ and $0$, respectively.
Lemma~\ref{lem:short_path} allows us to bound the (nonnegative) counters in vertices of
$\D'$ by $|V|^2+1$ and use them to make the strategy choose the right
successor in transitions of the type $(u,m,n)\bbtran{}[u,m,n,v]$ so that
$v$ is the successor of $u$ on the ``short path'' from Lemma~\ref{lem:short_path}.
If the strategy chooses a bad successor, the player gets ``punished''
in terms of not satisfying the $\MPONO{\D'}{r}$ objective
by entering a special vertex $\mathit{div}$ (for \emph{div}erge).
Indeed, if the counter of the steps overflows with the accumulated reward
from the last checkpoint being nonnegative, the play gets stuck
in $\mathit{div}$ and the objective $\MPONO{\D'}{r}$ is not satisfied.

\begin{lemma}\label{lem:computing_FD}
Let $s\in V$ be arbitrary.
The following is true.
\begin{enumerate}
\item Every MD strategy $\sigma'$ in $\D'$ satisfying 
$\calP(\MPS{\sigma'}{\D'}{r'}((s,1,0)))=1$ is decreasing.
\item For every MD strategy $\sigma$ in $\D$ there is a MD strategy 
$\sigma'$ in $\D'$ such that $\calP(\CNDS{\sigma'}{\D'}{r'}((s,1,0)))=1$
for every $s\in V$ such that $\calP(\CNDS{\sigma}{\D}{r}(s))=1$.
\end{enumerate}
\end{lemma}

\begin{proof}[of 1.]
First, observe that $\mathit{div}$ is not reachable from $(s,1,0)$.
Let $(u,n,m)$ be a state of $\D'\langle\sigma'\rangle$ reachable from
$(s,1,0)$. First, assume that $n>0$.
There is a path $w$ from $(u,n,m)$ to a state of the form
$(u',0,m')$, otherwise $\mathit{div}$ would have been reachable from
$(s,1,0)$. 
Let $k=\frac{|w|-1}{2}$.
For every $0\leq i\leq k$ we
denote $(v_i,n_i,m_i)=w(2i)$. Then
$n_0=n>0$ and
$n_i=n+\sum_{j=0}^{i-1} r(v_j)$
for $1\leq i\leq k$.
It follows that $n+r'(w)=n+\sum_{j=0}^{k-1} r(v_i)=n_k=0$.
This implies $r'(w)<0$. 

Now assume that $n=0$. Then $(u,n,m)\bbtran{} [u,1,0,v]$
and $[u,1,0,v]\bbtran{} (v,1+r(u),1)$. Denote
$w'=(u,n,m)[u,1,0,v]$. If $r(u)=-1$, then 
$r'(w'\cdot (v,1+r(u),1))=r(u)=-1$ and we are done. If $r(u)\geq 0$, then $1+r(u)>0$
and arguing as above we obtain a path $w$ from 
$(v,1+r(u),1)$ to some $(u',0,m')$ such that $1+r(u)+r'(w)=0$. However,
then $1+r'(w'w)=1+r(u)+r'(w)=0$ and $r'(w'w)=-1$.

Let $[u,n,m,v]$ be a state reachable from $(s,1,0)$.
Then $n>0$ and there is a transition $[u,n,m,v]\bbtran{} (v,n+r(u),m+1)$.
Arguing as above, we obtain that there is a path $w$ from
$(v,n+r(u),m+1)$ to some $(u',0,m')$ such that 
$n+r(u)+r'(w)=0$, which implies
that the $r'([u,n,m,v]\cdot w)=r(u)+r'(w)\leq-1$
and thus
$r'([u,n,m,v]\cdot w')=-1$
for some prefix $w'$ of $w$.
\end{proof}

\begin{proof}[of 2.]
For every $v\in V$ and $0\leq m\leq |V|^2$ we denote by $P(v,m)$ the set of all
paths in $\D\langle\sigma\rangle$ of length at most $|V|^2+1-m$ 
initiated in $v$. We denote
${val}(v,m)=\min\{r(w)\mid w\in P(v,m)\}$ and ${val}(v,|V|^2+1)=0$. 
For $m\leq |V|^2$ choose $\theta(v,m)\in V$ to be an arbitrary vertex $u$ such that
$v\tran{} u$ is a transition of $\D\langle\sigma\rangle$ and
\[
{val}(u,m+1)=\min\{{val}(u',m+1)\mid v\tran{} u'\text{ in }\D\langle\sigma\rangle\}
\]
Let us define a strategy $\sigma'$ as follows: 
\begin{itemize}
\item Let $(u,n,m)\in V'_N$.
  \begin{itemize}
  \item If $m=|V|^2+1$ and $n>0$, we put $\sigma'((u,n,m))={div}$
  \item If $m\leq |V|^2+1$ and $n=0$, we put $\sigma'((u,n,m))=[u,1,0,\theta(u,0)]$
  \item If $m<|V|^2+1$ and $n>0$, we put 
        $\sigma'((u,n,m))=[u,n,m,\theta(u,m)]$
  \end{itemize}
\item For every $[u,n,m,v]\in V'_N$, we put $\sigma'([u,n,m,v])=(v,n+r(u),m+1)$
\end{itemize}
Fix an arbitrary
$s\in V$ such that $\calP(\CNDS{\sigma}{\D}{r}(s))=1$.
Denote by $R$ the set of all states of the form
$(u,n,m)$ reachable from $(s,1,0)$.
We prove that
$n+{val}(u,m)\leq 0$
for all $(u,n,m)\in R$
by induction
on $m$. If $m=0$, then $n=1$ and Lemma~\ref{lem:short_path} implies
${val}(u,0)\leq -1$.

Consider $(u,n,m)\in R$ such that $m>0$.
Then
$(u',n',m')\tran{} [u',n'',m'',u]\tran{} (u,n,m)$ in $\D\langle\sigma\rangle$
for some $(u',n',m')\in R$.
Now either $n''=1$ and $m''=0$, or $n''=n'$ and $m''=m'$.
First, assume that $n''=1$ and $m''=0$. Then $n=1+r(u')$ and $m=1$. 
By 
Lemma~\ref{lem:short_path}, ${val}(u',0)\leq -1$ and thus by definition of
$\sigma'$, $1+r(u')+{val}(u,1)\leq 0$.
Now assume that $n''=n'$ and $m''=m'$. Then $n=n'+r(u')$ and $m=m'+1$.
By induction hypothesis,
$n'+{val}(u',m')\leq 0$, and thus by definition of
$\sigma'$, $n'+r(u')+{val}(u,m)\leq n'+{val}(u',m')\leq 0$.

This proves that if $(u,n,|V|^2+1)\in R$, then $n=0$. It follows that
$\mathit{div}$ is not reachable.
Given $[u,n,m,v]\in V'$, we define $\Theta([u,n,m,v])=u$.
Given $w\in \run_{\D'(\sigma')}((s,1,0))$, we define a run
$\Theta(w)\in \run_{\D(\sigma)}(s)$ by $\Theta(w)(k)=\Theta(w(2k+1))$.
We have that
$\Theta$ induces an isomorphism of the probability spaces
$\run_{\D'\langle\sigma'\rangle}((s,1,0))$ and $\run_{\D\langle\sigma\rangle}(s)$.
Indeed, it follows from the following three facts: First, ${div}$ is not reachable.
Second, if $u\in V_N$ and
$[u,n,m,v]\in R$, then $\sigma(u)=v$ and $[u,n,m,v]\tran{1}
(v,n'',m'')\tran{1} [v,n',m',v']$ in $\D'\langle\sigma'\rangle$ for some
$n'',m'',n',m',v'$. Third, if $u\in V_P$ and 
$[u,n,m,v]\in R$, then 
$[u,n,m,v]\tran{x}
(u',n'',m'')\tran{1} [u',n',m',v']$ in $\D'\langle\sigma'\rangle$ for some
$n'',m'',n',m',v'$ iff
$u\btran{} u'$ is assigned $x$ in $\D$.
Also, 
$w\in \CNDS{\sigma'}{\D'}{r'}((s,1,0))$ iff $\Theta(w)\in
\CNDS{\sigma}{\D}{r}(s)$. Thus
$\calP(\CNDS{\sigma}{\D}{r}(s))=
\calP(\CNDS{\sigma'}{\D'}{r'}((s,1,0)))=1$.
\end{proof}

So far we have that, for a fixed initial vertex $s\in V$, if $\SCN{\D}\neq\emptyset$ in $\D$
then $\SCN{\D'}=\SMP{\D'}{r}{\leq 0}\neq\emptyset$ in $\D'$.
It remains to prove the other implication. We do this in
two steps and we need the following notion:

\begin{definition}
A deterministic strategy $\sigma$ in $\D$ is said to be
\emph{finite-memory} (FD) if there is a deterministic finite automaton (DFA) 
$\A$ such that for every $wu\in V^*V$ the value of 
$\sigma(wu)$ depends only on $u$ and the current state $k$ of
$\A$ after reading $w$ (we write $\sigma(u,k)$ instead of $\sigma(wu)$).
\end{definition}

\begin{lemma}\label{lem:finite_memory_exist}
Given a MD strategy $\sigma'$ in $\D'$
there is a FD strategy $\sigma$ in $\D$ computable in polynomial time
such that $\calP(\CNDS{\sigma}{\D}{r}(s))=1$
for every $s\in V$ with $\calP(\CNDS{\sigma'}{\D'}{r'}((s,1,0)))=1$.
\end{lemma}

\begin{proof}
Let us define $\sigma$ as follows: Let $\A=(K,V,\zeta,k_{in})$ where
\begin{itemize}
\item $K$ consists of all vertices of $V'$ of the form
      $[u,n,m,v]$.
\item $\zeta$ is defined as follows:
      Let $[u,n,m,v]\in K$ and $u'\in V$. If $u\btran{} u'$, then
      we define $\zeta([u,n,m,v],u')$
      to be the unique vertex of the form $[u',n',m',v']$ satisfying
      $[u,n,m,v]\tran{} (u',n'',m'')\tran{} [u',n',m',v']$ in 
      $\D'\langle\sigma'\rangle$ for some $n''$ and $m''$. Otherwise, we define 
      $\zeta([u,n,m,v],u')$ to be an arbitrary state of $\A$.
\item Define $k_{in}=\sigma'((s,1,0))$.
\end{itemize}
For $u\in V_N$, we define $\sigma(u,[u,n,m,v])=v$. For $u'\not = u$ we define
$\sigma(u',[u,n,m,v])$ to be an arbitrary vertex $u''$ such that
$u'\btran{} u''$.

The rest is similar to the end of the proof of Lemma~\ref{lem:computing_FD}.
Given $[u,n,m,v]\in V'$, we define $\Theta([u,n,m,v])=u$.
Given $w\in \run_{\D'(\sigma')}((s,1,0))$, we define a run
$\Theta(w)\in \run_{\D(\sigma)}(s)$ by $\Theta(w)(k)=\Theta(w(2k+1))$.
Then $\Theta$ induces an isomorphism of the probability spaces
$\run_{\D'(\sigma')}((s,1,0))$ and 
$\run_{\D(\sigma)}(s)$.
Indeed, it follows from the following facts: 
First, ${div}$ is not reachable from $(s,1,0)$ in 
$\D'(\sigma')$. Second, if $[u,n,m,v]\in V'_N$, then
$[u,n,m,v]\tran{1} (v,n'',m'')\tran{1} [v,n',m',v']$ 
in $\D\langle\sigma'\rangle$ for some $n'',m'',n',m',v'$. Third,
for $[u,n,m,v]\in V'_P$, $[u,n,m,v]\tran{x} (u',n'',m'')\tran{1} [u',n',m',v']$ 
for some $n'',m'',n',m',v'$
iff the transition $u\btran{} v$ is assigned $x$ in $\D$.
Also, 
$w\in \CNDS{\sigma'}{\D'}{r'}((s,1,0))$ iff $\Theta(w)\in
\CNDS{\sigma}{\D}{r}(s)$. Thus
$\calP(\CNDS{\sigma}{\D}{r}(s))=
\calP(\CNDS{\sigma'}{\D'}{r'}((s,1,0)))=1$.
\end{proof}

\begin{remark}\label{rem:identify-FD}
Since the DFA $\A$ in the proof of Lemma~\ref{lem:finite_memory_exist}
effectively simulates the Markov Chain $\M$, we will simplify
the notation used in the procedure \FuncSty{CN-FD-to-MD}
by identifying the MD strategy for $\D'$ with its
associated FD strategy for $\D$.
\end{remark}

\begin{lemma}\label{lem:FD_to_MD}
Let $\sigma'$ be a FD strategy in $\D$.
Then the procedure \FuncSty{CN-FD-to-MD} computes
in time polynomial in the size of the DFA associated with $\sigma'$
a MD strategy $\sigma$
such that $\calP(\CNDSNO{\sigma}{\D}{r}(s))=1$
for every $s\in V$ with $\calP(\CNDS{\sigma'}{\D}{r}(s))=1$.
\end{lemma}

\begin{proof}
Denote $\A$ the DFA associated with $\sigma'$.
Further denote $K$ the set of its states, $\hat{k}$ its initial state,
$\zeta$ its transition function. Recall that the input alphabet
of such an automaton is $V$, the set of vertices of the MDP
$\D$.
We combine $\A$ with $\D$ and $\sigma'$ by means of parallel sequential
composition into a finite Markov chain $\M$.
More precisely, the set of vertices of $\M$ is the set
$V\times K$ and the transitions and probabilities
are defined as follows: For $u\in V_N$ and
$k\in K$ we put $(u,k)\tran{1} (u',k')$
if and only if $\sigma'(u,k)=u'$ and $k'=\zeta(k,u)$. For
$u\in V_P$ and $k\in K$ we put $(u,k)\tran{x} (u',k')$
if and only if $u\btran{} u'$ is assigned the probability $x$ and $k'=\zeta(k,u)$.
Given $(u,k)\in V\times K$, we denote the projection $\pi_1((u,k))=u$ and define $r'((u,k))=r(u)$.

The following procedure \FuncSty{CN-FD-to-MD} computes a sequence of Markov chains
$M_n$, $0\leq n\leq |V|$ with state spaces $V\times K$, transitions $\tran{}_{n}$ and
probabilities $\Prob_{n}$.
Then it extracts the strategy $\sigma$ from the last $M_n$, $n=|V|$.
For every $0\leq n\leq |V|$, let $\C_{n}$ be a union of all BSCCs of $M_{n}$ reachable from $(s,\hat{k})$.
We say that $u\in V$ is \emph{ambiguous in $\C_{n}$} if for at least two $k_1, k_2\in K$,
$k_1\neq k_2$, both $(u,k_1), (u,k_2) \in \C_{n}$.
For every $M_n$ an an initial vertex $(u,k)$ we define a random variable $R_{(u,k)}$ as follows:
given a run $w$ we set
$S=\{m>0 \mid \pi_1(w(m))=u\}$ and
put 
$$
R_{(u,k)}(w) = \begin{cases}
         r'(w{\downarrow}m) & S\neq\emptyset, m = \min S\\
         \bot               & S=\emptyset
         \end{cases}
$$
Since every $M_n$ is finite,
$R_{(u,k)}$ is almost surely defined whenever $(u,k)$ lies$s\in V$ with $\calP(\CNDS{\sigma'}{\D}{r}(s))=1$ in a BSCC,
and the expectation $ER_{(u,k)}$ is finite, see~\cite[Theorem~1.10.2]{Norris98}.

\begin{procedure}[t]
\caption{CN-FD-to-MD($\sigma'$) -- computing a MD CN-optimal strategy from a FD one.}\label{alg:fd-md}
\dontprintsemicolon
\KwData{The product Markov Chain $\M$ determined by the strategy $\sigma'$.}
\KwResult{Produce a CN-optimal MD strategy $\sigma$.}

$n\leftarrow 0$, $M_0\leftarrow \M$
\;
\nllabel{alg:fd-md-init}

\While{there are states ambiguous in $\C_n$}{
\nllabel{alg:fd-md-while}

Pick $(u_a,k)\in \C_n$, such that
$u_a$ is ambiguous in $\C_{n}$, $ER_{(u_a,k)}\leq 0$,
and $\calP_{(u_a,k)}(R<0)>0$.
\;
\nllabel{alg:fd-md-pick}

Compute $M_{n+1}$ from $M_n$ as follows:\\
Set $(v,k')\tran{x}_{n+1}(u_a,k)$ iff
$(v,k')\tran{x}_n(u_a,k'')$ for some $k''$.
\;
Set $(v,k)\tran{x}_{n+1}(u,k')$ iff
$(v,k)\tran{x}_n(u,k')$ for $u\neq u_a$.
\;
\nllabel{alg:fd-md-step}

$n\leftarrow n+1$\;
\nllabel{alg:fd-md-inc}
}%

$\C\leftarrow \{ u\in V \mid (u,k)\in \C_n\}$.
\;
\nllabel{alg:fd-md-C}

$(\varrho,-)$ $\leftarrow$ \FuncSty{Max-Reach}($\D$,$\C$)
\;
\nllabel{alg:fd-md-reach}

\lFor{$u \in V$}{\lIf{$u\notin \C$}{$\sigma(u) = \varrho(u)$} \lElse{$\sigma(u) = v$ where $\exists k,k' \in K:(u,k)\tran{}_n (v,k')$}} \;
\nllabel{alg:fd-md-sigma}

\Return{$\sigma$} \;
\end{procedure}

Procedure \FuncSty{CN-FD-to-MD} computes $\sigma$,
we first estimate its running time.
The while-loop on line~\ref{alg:fd-md-while} is executed
at most $|V|$-times because every state $(u_a,k)$ picked
in step \ref{alg:fd-md-pick} is no longer ambiguous in later
iterations. We show that the picking in step \ref{alg:fd-md-pick}
takes polynomial time.
First, due to, e.g.~\cite[XV.7]{Feller68}, $ER_{(u_a,k)}$ can be expressed as a unique solution of
a linear system of equations, computable in polynomial time.
So we can compute $ER_{(u_a,k)}$ in polynomial time and
check whether $ER_{(u_a,k)}\leq 0$. 
Second, the problem whether $\calP(R_{(u_a,k)}<0)>0$ is equivalent to the existence
of a negative weighted cycle in the BSCC containing $(u_a,k)$, which can be decided in
polynomial time using, e.g., the Bellman-Ford algorithm.
Time complexity of the procedure \FuncSty{Max-Reach}
on line~\ref{alg:fd-md-reach} has already been analyzed in Section~\ref{subsec:CN-to-qual}.

Let us prove correctness.
Fix some $s\in V$ with $\calP(\CNDS{\sigma'}{\D}{r}(s))=1$.
We prove by induction that for all $h$, $0\leq h\leq |V|$:
$\calP(\CNDONO{M_h}{r}((s,\hat{k})))=1$ in $M_h$.
For $h=0$ this is true by the choice of $s$.
Assume that the statement is true for some $h=n\in\Nset_0$,
we prove it for $h=n+1$.
First, we prove that if there is some $v$ ambiguous in $\C_n$, then there is
$(v,k)\in \C_n$ such that $ER_{(v,k)}\leq 0$ and $\calP(R_{(v,k)}<0)>0$. 
Let $C$ be a BSCC of $M_n$ reachable from $(s,\hat{k})$ and
containing at least two states from $\{v\}\times K$. Let us
denote $C^v\coloneqq(\{v\}\times K)\cap C=\{(v,k_1),\ldots,(v,k_{\ell})\}$.

We define sequences of random variables 
$I_0,I_1,I_2\ldots$ and $X^{i}_1,X^{i}_2,\ldots$ where
$i\in \{1,\ldots,\ell\}$
as follows: Let $w$ be a run
in $M_n$ initiated in some $(v,k_j)\in C^v$.
We define $I_0(w)=0$, and for all $j\geq 1$ we define 
$I_j(w)$ to be the least $m>I_{j-1}(w)$ such that $w(m)\in C^v$.
Let $i\in \{1,\ldots,\ell\}$
and let $m_1,m_2,m_3,\ldots$ be all indexes such that
$I_{m_j}(w)=(v,k_{i})$. 
We define 
$X^{i}_j(w)=r'(w{\downarrow}{I_{m_{j}+1}(w)})-
r'(w{\downarrow}{I_{m_{j}}(w)})$ 
the reward accumulated between the $j$-th visit to $(v,k_i)$
and next visit to $C^{v}$. 

Consider the Markov chain $M_n$.
Observe that $EX^i_1$ is independent of the actual
initial vertex $(v,k_j)\in C^v$ and
that $EX^i_1=ER_{(v,k_i)}$.
Also, for a fixed $i,1\leq i\leq \ell$, the variables $X^{i}_j$, $j\geq 1$ 
are independent and identically distributed.
We claim that $EX^{i}_1\leq 0$ for some $i$ and 
$\calP(X^i_1<0)>0$. Assume, to the contrary, that
there is no such $i$ and let us denote $B=\{i\mid EX^i_1>0\}$.
The variables $X^{i}_j$ generate $\ell$ random walks of the form 
$S^i_1,S^i_2,\ldots$
by $S^{i}_{n}=\sum_{j=1}^n X^{i}_j$. For every $i\in B$ the
walk $S^i_j$ drifts almost surely to $\infty$,
by, e.g., \cite[Theorem 8.3.4]{Chung01}.
On the other hand, for every $i\in \{1,\ldots,\ell\}\smallsetminus B$
the walk never reaches values smaller than a fixed number.
Since for almost all runs starting in some $(v,k_j)\in C^v$
we have $\liminf_{n\to\infty}r(w{\downarrow}n)=\liminf_{n\to\infty}\sum_{i=1}^\ell S^i_n$,
it follows that in $M_n$: $\calP(\CNDONO{M_n}{r}((v,k_j)))=0$, 
and hence $\calP(\CNDONO{M_n}{r}((s,\hat{k})))<1$, a contradiction.
	
Now we prove that in $M_{n+1}$: $\calP_{(s,\hat{k})}(CN)=1$.
Assume that $(u_a,k)$ is the state selected in step \ref{alg:fd-md-pick}.
Then the expected value $ER_{(u_a,k)}$ is the same in both $M_{n+1}$
and $M_n$ and thus not positive.
Further $\calP(R_{(u_a,k)}<0)>0$ in $M_{n+1}$
and no states of the form $(u_a,k')$ where $k'\neq k$ are reachable from $(u_a,k)$ in $M_{n+1}$.
By Lemma~\ref{lem:mp_nonpos},
$\calP(\CNDONO{M_{n+1}}{r}((u_a,k)))=1$ in $M_{n+1}$.
Let $A$ be the set of all
runs of $\run_{M_n}((s,\hat{k}))$ not reaching $(u_a,k)$.
Clearly, the probability of $A$ is the same in $M_n$ as in $M_{n+1}$.
Hence, $\calP(\CNDONO{M_{n+1}}{r}((s,\hat{k})))=1$ in $M_{n+1}$.

Finally, note that for every $n$ the Markov chain $M_n$ has the following
properties: 
\begin{itemize}

\item For each $(u,k)\in V_P\times K$, if $(u,k)\tran{x} (v,k')$, then
      $u\btran{x} v$.

\item For each $(u,k)\in V_N\times K$, if $(u,k)\tran{x} (v,k')$, then
      $x=1$ and $u\btran{} v$.

\end{itemize}
Hence, in step~\ref{alg:fd-md-C}
the set $\C$ is reachable from $s$ with probability $1$
using a suitable MD strategy $\varrho$, line~\ref{alg:fd-md-reach}.
Consequently the strategy $\sigma$ for $\D$ is well-defined
(line~\ref{alg:fd-md-sigma}) and satisfies
$\calP(\CNDSNO{\D}{r}{\sigma}(s))=1$.
\end{proof}

The correctness of the reduction represented by the procedure
\FuncSty{Qual-CN} follows from
Lemma~\ref{lem:computing_FD},
Lemma~\ref{lem:finite_memory_exist},
Remark~\ref{rem:identify-FD}, and
Lemma~\ref{lem:FD_to_MD}.

\subsection{Correctness of \FuncSty{Qual-MP}}\label{subsec:corr-MP}
Denote $W=\{s \in V \mid \exists \text{a MD strategy } \sigma: \calP(\MPSNO{\sigma}{\D}{r}(s))=1\}$.
In this section we prove that the set $A$ and MD strategy $\sigma$ computed by
the procedure \FuncSty{Qual-MP} satisfy:
$\forall s \in W: \calP(\MPSNO{\sigma}{\D}{r})=1$
and $W=A$.

Choose an arbitrary $s \in V$.
Let $\sigma$ be a MD strategy. 
Let us denote $\bscc{\D\langle\sigma\rangle}$
the set of all BSCCs of $\D\langle\sigma\rangle$ reachable from $s$.
By standard arguments from the theory of Markov chains (see~e.g.~\cite[Section~1.5]{Norris98}),
$\sum_{C\in \bscc{\D\langle\sigma\rangle}} 
\calP({Reach}^{\sigma}_{C}(s))=1$.
Recall also the random variable $\rwV{\sigma}{r}{s}$ defined in Section~\ref{subsec:solve-MP}.
In particular recall that \cite[Theorem~1.10.2]{Norris98} implies that for almost all runs $w$
$\rwV{\sigma}{r}{s}(w)=\lim_{n\to \infty} \frac{r({w}\downarrow{n})}{n}$.
Moreover, using \cite[Theorem~1.10.2]{Norris98} again,
for every $C\in\bscc{\D\langle\sigma\rangle}$
there is a constant $a_C\in \Rset$ such that $\rwV{\sigma}{r}{s}=a_C$
almost surely on the condition of hitting $C$.
Thus for the expected value we have
\[
E\rwV{\sigma}{r}{s}=\sum_{C\in \bscc{\D\langle\sigma\rangle}} 
a_C\cdot\calP({Reach}^{\sigma}_{C}(s))
\]
We prove that there is a MD strategy $\varrho$ computable in polynomial time
such that $E\rwV{\varrho}{r}{s}=\min_\sigma E\rwV{\sigma}{r}{s}$
where the minimum is taken over MD strategies.

Let $\sigma$ be a MD strategy. 
We define a sequence of random variables 
$\rwiV{1}{\sigma}{r}{s},\rwiV{2}{\sigma}{r}{s},\ldots$ 
such that
$\rwiV{n}{\sigma}{r}{s}=r(w{\downarrow} n)$
for every run $w\in \run_{\D\langle\sigma\rangle}(u_0)$ 
and every $n\geq 1$.
Let us denote $E\rwiV{n}{\sigma}{r}{s}$ 
the expected value of $\rwiV{n}{\sigma}{r}{s}$
(i.e. $E\rwiV{n}{\sigma}{r}{s}=\sum_{i=-n}^n i\cdot \calP(\rwiV{n}{\sigma}{r}{s}=i)$).

Note that
\[
\frac{E\rwiV{n}{\sigma}{r}{s}}{n}=\frac{\sum_{i=-n}^n i\cdot \calP(\rwiV{n}{\sigma}{r}{s}=i)}{n}=
\sum_{i=-n}^n \frac{i}{n}\cdot\calP(\frac{\rwiV{n}{\sigma}{r}{s}}{n}=\frac{i}{n})=
E\frac{\rwiV{n}{\sigma}{r}{s}}{n}
\]
and that $|\frac{\rwiV{n}{\sigma}{r}{s}}{n}|\leq 1$. 
Hence by the dominated convergence theorem (see e.g.~\cite[Theorem~16.4]{Billing95})
\[
\lim_{n\to \infty} \frac{E\rwiV{n}{\sigma}{r}{s}}{n}=
\lim_{n\to \infty} E\frac{\rwiV{n}{\sigma}{r}{s}}{n}=E\rwV{\sigma}{r}{s}
\]
Using either~\cite[Theorem~2.9.4]{FilarVrieze}, or~\cite[Theorem~9.3.8]{Puterman94}, and 
a P-time algorithm for linear programming, one can construct
a polynomial time algorithm which computes
a MD strategy $\varrho$ such that (taking the minima over MD strategies)
\[
E\rwV{\varrho}{r}{s}=\lim_{n\to \infty} \frac{E\rwiV{n}{\varrho}{r}{s}}{n}=
\min_\sigma \lim_{n\to \infty} \frac{E\rwiV{n}{\sigma}{r}{s}}{n}=
\min_\sigma E\rwV{\sigma}{r}{s}
\]
and also computes the value 
$E\rwV{\varrho}{r}{s}$.

In the proof of correctness and the complexity estimates of \FuncSty{Qual-MP} we will
denote $r_i$, $T_i$, and $A_i$ the reward represented by $\hat{r}$, the content of the set $T$,
and the set $A$, respectively, before
the $i$-th iteration of the while-loop,
in particular $r_0=r$, $T_0=\emptyset$, and $A_0=\emptyset$.
We also denote $\varrho_i$ the strategy $\varrho$ from line~\ref{proc:mp-qual-varrho}
computed in the $i$-th iteration of the while-loop.

Choose some $s\in W$ so that there is a strategy
$\bar{\sigma}$ such that $\calP(\MPSNO{\bar{\sigma}}{\D}{r}(s))=1$, i.e.,
$\rwV{\bar\sigma}{r}{s}(w)\leq 0$ almost surely.
Given $i$, we define a MD strategy $\sigma^i$ such that for every $u\in V$
      \[
      \sigma^i(u) = \begin{cases}
                  v & (u,v)\in T_{i} \\
                  \bar{\sigma}(u) & \text{ otherwise.}
                  \end{cases}
      \]      
The algorithm keeps the following invariants: 
\begin{enumerate}

\item[(a)]
$\rwV{\sigma^i}{r_i}{s}(w)\leq 0$ and $\rwV{\sigma^i}{r}{s}(w)\leq 0$ almost surely.

\item[(b)] For every $u\in V\smallsetminus A_i$ and every strategy $\sigma$ in $\D$,
the probability of reaching $A_i$ from $u$ is strictly less than $1$.
There is no path from any state of $A_i$ to 
$V\smallsetminus A_i$ in $\D\langle\sigma^i\rangle$.

\item[(c)] $A$ and $V_?$ are disjoint.
\end{enumerate}
The invariant (c) follows by an easy induction from lines~\ref{proc:mp-qual-init}
and \ref{proc:mp-qual-s-back}.

Clearly, the invariant (a) implies that on line~\ref{proc:mp-qual-varrho} the strategy $\varrho$ always
exists. We prove that a BSCC $C$ from line~\ref{proc:mp-qual-C} exists. 
Note that by the invariant (b), for all 
$C\in \bscc{\D\langle\sigma\rangle}$ either $C\cap A_i=\emptyset$,
or $C\subseteq A_i$, and there must be at least one
$C$ such that $C\cap A_i=\emptyset$, otherwise $s$ could not have been in $A_i$,
contradicting line~\ref{proc:mp-qual-s} and the invariant (c).
Also there are numbers $a_{C,i}$ for every $C\bscc{\D\langle\varrho_i\rangle}$ such that
$\rwV{\varrho_i}{r_i}{s}=a_{C,i}$ almost surely on the condition
of hitting $C$, and
\[
E\rwV{\varrho_i}{r_i}{s}=\sum_{C\in \bscc{\D\langle\varrho_i\rangle}} 
a_{C,i}\cdot\calP({Reach}^{\varrho_i}_{C}(s))
\]
However, all $C\in \bscc{\D\langle\varrho_i\rangle}$ such that
$C\subseteq A_i$ satisfy $a_{C,i}=0$. Hence, there must be at least
one $C_{wit}\in \bscc{\D\langle\varrho_i\rangle}$ such that
$C_{wit}\cap A_i=\emptyset$ and
$a_{C_{wit},i}\leq 0$.

Now every $D\in\bscc{\D\langle\sigma^{i+1}\rangle}$ satisfies
either $D=C_{wit}\subseteq A_{i+1}\smallsetminus A_i$, or
$D\subseteq (V\smallsetminus A_{i+1})\cup A_i$ and
$D\in \bscc{\D\langle\sigma^{i}\rangle}$.
Moreover, transitions
between states of $C_{wit}$ in $\D\langle\sigma^{i+1}\rangle$ coincide with
transitions between states of $C_{wit}$ in $\D\langle\sigma_{i}\rangle$.
Also, transitions between states of every $D\neq C_{wit}$ in $\D\langle\sigma^{i+1}\rangle$
coincide with transitions between states of $D$ in $\D\langle\sigma^{i}\rangle$.

Then almost all $w\in {Reach}^{\sigma^{i+1}}_{C_{wit}}(s)$
satisfy $\rwV{\sigma^{i+1}}{r_{i+1}}{s}(w)\leq 0$ because
$r_{i+1}$ assigns $0$ to all states of $C_{wit}$. Also, 
almost all $w\in {Reach}^{\sigma^{i+1}}_{D}(s)$
where $D\neq C_{wit}$
satisfy $\rwV{\sigma^{i+1}}{r_{i+1}}{s}(w)=
\rwV{\sigma^{i}}{r_{i}}{s}(w)\leq 0$ due to the invariant (a) for $i$. It follows
that
$\rwV{\sigma^{i+1}}{r_{i+1}}{s}(w)\leq 0$
for almost all runs $w\in\run_{\D\langle\sigma^{i+1}\rangle}(s)$.

Moreover,
almost all $w\in  {Reach}^{\sigma^{i+1}}_{C_{wit}}(s)$
satisfy $\rwV{\sigma^{i+1}}{r}{s}(w)\leq 0$ because $r_i$ coincides with $r$
on $C_{wit}$ and almost all runs $w\in 
 {Reach}^{\sigma_{i}}_{C_{wit}}(s)$
satisfy 
$\rwV{\sigma_{i}}{r_i}{s}(w)\leq 0$.
Also, 
almost all $w\in  {Reach}^{\sigma^{i+1}}_{D}(s)$
where $D\neq C_{wit}$
satisfy $\rwV{\sigma^{i+1}}{r}{s}(w)=
\rwV{\sigma^{i}}{r}{s}(w)\leq 0$ due to the invariant (a) for $i$.
Hence, for almost all runs $w\in\run_{\D\langle\sigma^{i+1}\rangle}(s)$
we have $\rwV{\sigma^{i+1}}{r}{s}(w)\leq 0$. 

It follows that the invariant (a) is preserved.
The invariant (b) is preserved due to computation of $\tau$ on line~\ref{proc:mp-qual-tau},
$A'$ on line~\ref{proc:mp-qual-Ap}
and update of $A$ in line~\ref{proc:mp-qual-A-next}.
Finally, the strategy $\sigma$ defined on line~\ref{proc:mp-qual-sigma}
has the desired properties
because it coincides with $\sigma^{i+1}$ on all reachable states, and
$\sigma^{i+1}$ satisfies the invariant (a).
This also implies that the vertex $s$ was put into $A'$
on line~\ref{proc:mp-qual-Ap} and consequently to $A$
on line~\ref{proc:mp-qual-A-next} in some iteration of the while-loop.
Thus $W \subseteq A$.
Since by arguments similar as above we can show that
for every $s\in A$ we have
$\calP(\MPSNO{\sigma}{\D}{r}(s))=1$ the correctness
is proved.

Let us now consider the complexity.
By~\cite[Theorem~1.10.2]{Norris98},
for every $C\in \bscc{\D\langle\varrho_i\rangle}$
the is a constant $a_{C,i}\in \Rset$ defined above
is equal to $\sum_{u\in C} \mu(u)\cdot r_i(u)$,
where $\mu$ is the invariant distribution for $C$ (note that $C$ can be
considered as a standalone irreducible Markov chain within $\D\langle\varrho_i\rangle$),
which is a unique solution
of a system of linear equations, and thus computable in polynomial time. 
Hence, a suitable BSCC satisfying the conditions
from line~\ref{proc:mp-qual-C} can be computed in polynomial time.
In Section~\ref{subsec:solve-MP} we already showed that
the strategy $\varrho$ from line~\ref{proc:mp-qual-varrho}
can be found in polynomial time.
In Section~\ref{subsec:CN-to-qual} we showed that also finding
the strategy $\tau$ on line~\ref{proc:mp-qual-tau} can be done in polynomial time.
Other steps can be clearly taken in polynomial time.
Since the set $A$ grows with every iteration of the while-loop
by at least one vertex, the loop itself is executed at most $|V|$-times.
Thus the procedure \FuncSty{Qual-MP} runs in polynomial time.

%% file: app-bound.tex
\section{Proofs of \protect{Section~\ref{sec-boundary}}}
\label{app-boundary}

\newcommand{\BH}{\textbf{BH}}
\newcommand{\DP}{\textbf{DP}}

\noindent
For the rest of this section, we fix an OC-MDP
$\A = (Q,\delta^{=0},\delta^{>0},(Q_N,Q_P),P^{=0},P^{>0})$ and 
a non-empty set $F \subseteq Q$ of final states. We assume 
(without restrictions) that for each $q \in F$, the configuration 
$q(0)$ has only one outgoing transition $q(0) \ctran{} q(0)$.
We also use $N$ to denote $2^{|Q|}$.

Obviously, $\OptValOne^{\NT} \subseteq \ValOne^{\NT}$ and
$\OptValOne^{\ST} \subseteq \ValOne^{\ST}$, but it is not immediately
clear whether the inclusions are proper. 
As we shall see, the sets $\OptValOne^{\NT}$, $\ValOne^{\NT}$, and
$\OptValOne^{\ST}$
have a regular structure which can be captured by finite state 
automata, and optimal strategies are either counter-oblivious
or counter-regular.

\begin{definition}[regular sets of configurations, counter-regular strategies]
  An \mbox{\emph{$\A$-automaton}} is a pair $(M,\varrho)$ where
  $M = (C,\{a\},\gamma,F)$ is a deterministic finite-state automaton
  and $\varrho : Q \rightarrow C$ a mapping.
  A set of configurations of~$\A$ \emph{recognized} by $(M,f)$ consists
  of all $p(i) \in Q \times \Nseto$ such that $M$ accepts the word 
  $a^i$ from the initial state $\varrho(p)$.  
  A set of configurations is \emph{regular} if it is recognized
  by some $\A$-automaton.

  A MD strategy $\sigma$ is \emph{counter-regular} if there is an 
  $\A$-automaton $(M,\varrho)$ and a function 
  $f : Q \times C \rightarrow \delta^{>0}$, where $C$ is the set of states
  of $M$, such that for all $p(i) \in Q \times \Nset$ we have that 
  $\sigma(p(i)) = f(p,q)$, where $q \in C$ is the state 
  entered from $\varrho(p)$ after reading the word $a^i$.
\end{definition}
We start by proving the results about $\NT$ objectives. 

\begin{reftheorem}{Theorem}{\ref{thm-NT}}
  The sets $\ValOne^{\NT}$ and $\OptValOne^{\NT}$ are equal. Moreover, 
  given a OC-MDP, $\A$, and a configuration $q(i)$ of $\A$,
  we can decide in polynomial time whether $q(i) \in \ValOne^{NT}$. 
  Furthermore, there is a $\SMD$ 
  strategy $\sigma$ constructible in polynomial time which is 
  optimal in every configuration of $\ValOne^{\NT} = \OptValOne^{\NT}$.
\end{reftheorem}
\begin{proof}
We start by showing that for all $i\geq |Q|$ and all $p \in Q$
such that $p(i) \in \ValOne^{\NT}$ we have that
\begin{equation}\label{eq:CN-is-NT}
1=
\sup_{\tau\in \HR}\calP(\NT^\tau(p(i)))
=
\sup_{\tau\in \HR}\calP(\CNCS{\tau}{\A}(p(i)))
\end{equation}
Let us fix some $p(i) \in Q \times \Nset_0$ where $i \geq |Q|$. 
Consider an arbitrary HR strategy $\tau$ for $\Dz_\A$.
For every $0 \leq j \leq i$, we define the set $U_j^\tau \subseteq Q$
which consists of all $q \in Q$ such that
with probability $>0$ a run from $p(i)$ under $\tau$
visits $q(j)$ before visiting any other configuration
$s(k)$ with $k\leq j$.
Consider further an arbitrary infinite sequence $\varepsilon_1, \varepsilon_2,
\ldots$ of positive reals where $\lim_{n\to \infty}\varepsilon_n=0$, and an
infinite
sequence of strategies $\sigma_1, \sigma_2,\ldots$ such that
$\calP(\NT^{\sigma_j}(p(i))) \geq 1-\varepsilon_j$ for all $j$.  Since
there are only finitely many collections of $i+1$ subsets of $Q$,
there are subsequences $\varepsilon_{d_1}, \varepsilon_{d_2}, \ldots$ and
$\sigma_{d_1}, \sigma_{d_2},\ldots$, and a collection
$U_0,\ldots,U_i\subseteq Q$ such that $\lim_{n\to
  \infty}\varepsilon_{d_n}=0$, $\calP(\NT^{\sigma_{d_j}}(p(i)))\geq
1-\varepsilon_{d_j}$ for all $j$, and moreover $U_k=U_k^{\sigma_{d_j}}$
for all $j$ and all $0\leq k \leq i$.

Since $i+1>|Q|$, there must be some $k$, where $0 \leq k \leq i$, 
such that $U_k \subseteq \bigcup_{i\geq j > k} U_j$.
Thus, for every $q\in U_k$ and $l\in\Zset$ the strategies $\sigma_{d_j}$, $j\geq 1$ induce
strategies in $\Dn_\A$ for reaching $U_k\times\{l-1,l-2,\ldots\}$ from
$q(l)$ with a probability arbitrarily close to $1$.
This allows us to construct strategies for satisfying
$\CNCO{\A}$
with probability arbitrarily close to $1$
from every $q(l)$, $q\in U_k$, $l\in\Zset$.
Indeed, for an arbitrary $\delta>0$ consider the sequence
$\{\delta_j\}_{j=1}^\infty$, where $\delta_j = \delta\cdot 2^{-j}$.
For every $w\in(Q\times\Zset)^+$ which starts with some $q(l) \in U_k\times\Zset$
we denote \emph{min-step} every index $j$ such that 
\begin{itemize}
\item $w(j)=q(m)$ for some $q\in U_k$, $m\in\Zset$,
\item for all $h$ such that $0\leq h < j$ we have that if $w(h)=q(m')$,
 then $q\notin U_k$ or $m'>m$.
\end{itemize}
We define a strategy $\tau$ by setting $\tau(w)=\tau_j(w')$ where
$j$ is the number of min-steps in $w$,
$w'=w(m)\cdots w(|w|-1)$ with $m$ being the last min-step,
and $\tau_j$ is a $\delta_j$-optimal strategy for
satisfying $\CNCO{\A}$ from $w(m)$.
It follows that $\calP(\CNCS{\tau}{\A}(q(l)))\geq\prod_{j=1}^\infty 1-\delta_j\geq1-\delta$.
Since the strategies $\sigma_{d_j}$ also induce strategies for
reaching $U_k\times\Zset$ from $p(i)$ with probability arbitrarily close to $1$,
we proved (\ref{eq:CN-is-NT}).

By applying Theorem~\ref{thm:optimal}, we can conclude that our 
theorem is true for all configurations
of the form $p(i)$ with $p\in Q$, $i\geq|Q|$, since an optimal $\SMD$
 strategy for $\CNCO{\A}$
induces directly an optimal $\SMD$ strategy in $\Dz_\A$ for $\NT$.
Let us denote this strategy by~$\sigma$.

Consider now the case $p(i)$ when $i<|Q|$.
Let 
\[
  A = \ValOne^{\NT}\cap \{q(j)\mid q\in Q, j\geq|Q|\}
\]
Consider a finite MDP $\D$ with vertices $Q\times\{0,1,\ldots,|Q|\}$
such that for all $q\in Q$ the vertices $q(|Q|)$ are stochastic
with only one transition $q(|Q|)\tran{1}q(|Q|)$ and the rest
is just restriction of transitions and probabilities from $\Dz_\A$.
Then the following is equivalent due to standard results for finite MDP
(see e.g. \cite{CY98}):
\begin{itemize}
\item $p(i)\in \ValOne^{\NT}$
\item There are strategies in $\Dz_\A$ for reaching $A\cup (Q\times\{0\})$
from $p(i)$
with probability arbitrarily close to $1$.
\item There are strategies in $\D$ for reaching $(A\cap(Q\times\{|Q|\}))\cup (Q\times\{0\})$
from $p(i)$
with probability arbitrarily close to $1$.
\item There is a MD strategy $\tau$ in $\D$ computable in polynomial time for reaching $(A\cap(Q\times\{|Q|\}))\cup (Q\times\{0\})$
from $p(i)$
with probability $1$.
\item $p(i)\in \OptValOne^{\NT}$ with the witnessing strategy being $\tau$ extended with $\sigma$
for configurations $q(m)$, $m\geq|Q|$.
\end{itemize}

We have already defined a CMD strategy $\sigma$ such that
$\calP(\NT^\sigma(p(i)))=1$ for all $i\in\Nset$ and $p$ such that
$\{p\}\times\Nset\subseteq \ValOne^{\NT}$ (call these $p$
\emph{safe}).  To finish the proof of our theorem, it remains to
redefine $\sigma$ for configurations $p(i)$, $p\in Q$, $i<|Q|$ such
that $p(i)\in\ValOne^{\NT}$ but $p(|Q|)\notin\ValOne^{\NT}$ (call
these $p$ \emph{unsafe}).
Note that due to (\ref{eq:CN-is-NT}) every $p\in Q$ is either safe
or unsafe.
For every unsafe $p$ there is some
$i_p<|Q|$ such that $p(i)\in\ValOne^{\NT}$ iff $i\leq i_p$.  Take the
MD strategy $\tau$ such that $\calP(\NT^\tau(p(i_p)))=1$.  Note that
this strategy can be chosen one for all such $p(i_p)$.  We now
redefine the CMD strategy $\sigma$ by redefining its selector $f$:
$f(p)$ is the rule generating the transition chosen by $\tau$ in
$p(i_p)$.  Since no configuration with an unsafe state is reached from
a configuration with a safe state under $\sigma$ this does not
influence the property that $\calP(\NT^\sigma(p(i)))=1$ for all safe
$p$.  Moreover from the definition of $f$ and the choice of $i_p$,
almost all runs from $p(i)$, $i\leq i_p$ under $\sigma$ either visit a
configuration with a safe state or a configuration from $Q\times\{0\}$
or $q(i_q+i-i_p)$ with $q$ unsafe. Thus by double induction, first on
$|Q|-i_p$ then on $i$, for all unsafe $p$ and $i\leq i_p$ we have
$\calP(\NT^\sigma(p(i)))=1$.

Since $\ValOne^{\NT}= \{(q,i)\mid q\text{ is safe}, i\in\Nset\} \cup
\{(q,i)\mid q\text{ is unsafe},i\leq i_q\}$, we have proved the theorem.
\qed
\end{proof}

\begin{remark}\itshape
\label{rem-NT-structure}
Let $I = \{p \in Q \mid p(i) \in \ValOne^{\NT} \mbox{ for all } i \in \Nset\}$.
Then 
\begin{itemize}
\item for every $q \in I \cap Q_P$ we have that if $(q,c,q') \in \delta^{>0}$, 
  then $q' \in I$;
\item if $q \in I \cap Q_N$, then there is $(q,c,q') \in \delta^{>0}$ such 
  that $q' \in I$.
\end{itemize}
This means that we can define a OC-MDP $\A_I$ obtained from $\A$ by 
\begin{itemize}
\item restricting the set of control states to $I$; 
\item restricting the set of positive rules to the rules of the
  form $(q,c,q')$ where $q,q' \in I$ and the probability assignment 
  is preserved;
\item redefining the set of zero rules to $\{(q,0,q) \mid q \in I\}$.
\end{itemize}
It follows from the proof of Theorem~\ref{thm-NT} that for every 
configuration $p(i)$ of $\A_I$ we have that 
$\Val^{\CN}(p(i)) = \Val^{\NT}(p(i)) = 1$.
\end{remark}

Now we give the promised example which demonstrates that
the inclusion $\OptValOne^{\ST} \subseteq \ValOne^{\ST}$ is
proper.
Consider the OC-MDP $\hat{\A}$
of the following figure (we draw directly the associated 
MDP $\Dz_{\hat{\A}}$):

\begin{center}
\tikzstyle{n}=[rectangle,thick,draw,minimum size=1.5ex,inner sep=0em]
\tikzstyle{p}=[circle,thick,draw,minimum size=1.5ex,inner sep=0em]
\tikzstyle{t}=[thick,draw,->,>=stealth]
\tikzstyle{loop left}=[t,
  to path={.. controls +(140:.7) and +(220:.7) .. (\tikztotarget) \tikztonodes}]
\newcounter{ll}\setcounter{ll}{0}
\begin{tikzpicture}[x=.8cm,y=.5cm,font=\footnotesize]
\foreach \i in {0,1,2,3,4,5,6}
  {\foreach \q in {0,1} 
      {\node (s\q\i) at (\i,\q) [p] {};}
   \foreach \q in {2}
      {\node (s\q\i) at (\i,\q) [n] {};}
   \node at (\i,-.7) {$\i$};
  }
\node at (-.7,2) {$p$};
\node at (-.7,1) {$r$};
\node at (-.7,0) {$s$};
\node (s07) at (7,0) [p,draw=none] {};
\node (s17) at (7,1) [p,draw=none] {};
\node (s27) at (7,2) [n,draw=none] {};
\node at (0,0) [p,fill=black] {};
\foreach \i in {1,2,3,4,5,6}
  {\setcounter{ll}{\i}\addtocounter{ll}{1}
   \draw [t] (s0\arabic{ll}) to (s0\i);
   \draw [t] (s1\i) to (s0\i);
   \draw [t] (s1\arabic{ll}) to (s1\i);
   \draw [t] (s2\i) to (s2\arabic{ll});
   \draw [t] (s2\i) to (s1\i);
  }
\draw [t] (s01) to (s00);
\draw [t] (s11) to (s10);
\draw [t] (s20) to (s21);
\draw [t] (s00) to [loop left] (s00);
\draw [t] (s10) to [loop left] (s10);
\draw [thick,dotted] (s07) -- +(1,0);
\draw [thick,dotted] (s17) -- +(1,0);
\draw [thick,dotted] (s27) -- +(1,0);
\end{tikzpicture}
\end{center}

\noindent
The control state $p$ is non-deterministic, and the other two control 
states are stochastic. The probability distributions are always uniform, 
and the only final control state is $s$. Now observe that   
$\OptValOne^{\ST} = \{s(i) \mid i\in\Nset_0\}$, while $\ValOne^{\ST}$ consists of all
$p(i)$, $s(i)$, $i \in \Nset_0$. To see this, let us
fix an arbitrarily small $\varepsilon > 0$, and choose some
$c \in \Nset_0$ such that $\frac{1}{2^c} < \frac{\varepsilon}{2}$.
We define a MD strategy $\sigma_\varepsilon$ by
$\sigma_\varepsilon(p(k)) = p(k+1)$ if $k<c$,
and $\sigma_\varepsilon(p(k)) = r(k)$ if $k\geq c$.
Now it is easy to check that $\calP(\ST^{\sigma_\varepsilon}(v)) \geq 1- \frac{1}{2^c} > 1-\varepsilon$
for every $v$ of the form $p(i)$, or $s(i)$.
On the other hand, there is no strategy $\sigma$ such that 
$\calP(\ST^{\sigma}(p(i))) = 1$ for any $i\in\Nset_0$
because every strategy which makes the 
probability of reaching $s(0)$ from $p(i)$ positive inevitably makes
the probability of reaching $r(0)$ positive as well.

Note that the strategy $\sigma_\eps$ from the above example is in fact
both MD and FD strategy (see the definition after Lemma~\ref{lem:computing_FD}),
i.e. finitely representable by a deterministic finite automaton.
This is always the case for strategies approximating
the $\Val^\ST$ up to some fixed $\varepsilon>0$.
This is because if some strategy $\sigma$ satisfies
$\calP(\ST^\sigma(v))\geq\Val^\ST(v)-\varepsilon/2$ then
there is some $n\in\Nset$ such that the probability of
runs from $\ST^\sigma(v)$ not longer than $n$
is at least $\Val^\ST(v)-\varepsilon$.
On these runs only finitely many configurations appear
and thus the choices of $\sigma$ in these configurations
can be kept in a finite memory of a finite automaton.
Thus the strategy $\sigma$ can be replaced by a FD
strategy $\sigma'$ copying the choices of $\sigma$ until
the $n$-th step. It follows that
$\calP(\ST^{\sigma'}(v))\geq\Val^\ST(v)-\varepsilon$.

Now we present an exponential-time algorithm which 
computes an $\A$-automaton recognizing the set 
$\OptValOne^{\ST}$, and we also show that there is a counter-regular
strategy $\sigma$ constructible in exponential time which is optimal
in the configurations of $\OptValOne^{\ST}$. We also give a lower complexity 
bound and show that deciding the membership to $\OptValOne^{\ST}$ 
is $\PSPACE$-hard,
and the membership to $\ValOne^{\ST}$ is hard for the Boolean hierarchy over $\NP$
(note this hierarchy subsumes both $\NP$ and $\coNP$). 
We did not manage to provide analogous results for $\ValOne^{\ST}$, and
we leave this problem as an open challenge for future work (the above example
gives a taste of issues that must be resolved to obtain a solution).

\noindent
To prove Theorem~\ref{thm-optvalone}, we need to formulate 
several auxiliary observations. For every $i \in \Nset_0$, let
\begin{itemize}
\item $\Black_i = \{p(i) \in Q\times\Nset_0 \mid R_i(p) = b\}$
\item $\White_i = \{p(i) \in Q\times\Nset_0 \mid R_i(p) = w\}$
\end{itemize}
Further, let $\White = \bigcup_{i \in \Nset_0} \White_i$.

\begin{lemma}
\label{lem-reach}
  There is a MD strategy
  $\sigma$ such that for all $0 \leq j < i$
  and all $p(i) \in \Black_i$ we have 
  that $\calP(\Reach_{\Black_{j}}^\sigma(p(i))) = 1$ and
  \mbox{$\calP(\Reach_{\White}^\sigma(p(i))) = 0$}.
\end{lemma}
\begin{proof}
  It is known that for every finitely-branching MDP
  $\D = (V,{\btran{}},(V_N,V_P),\Prob)$, every set $T \subseteq V$ of target 
  vertices, and every initial vertex $v \in V$, if there is 
  \emph{some} (i.e., HR)
  strategy $\pi_v$ such that $\calP(\Reach_T^{\pi_v}(v)) =1$, then there is
  also a MD strategy $\sigma_v$ with this property
  (see, e.g., Theorem 7.2.11 of \cite{Puterman94}, which
   applies to more general non-negative bounded 
    total expected reward objectives). 
  The individual 
  MD strategies $\sigma_v$ can be easily combined into a single 
  MD strategy $\sigma$. Since $\Dz_{\A}$ is finitely-branching, we can
  apply this generic result and conclude that there is a MD strategy
  $\sigma$ such that $\calP(\ST^\sigma(p(i))) = 1$ for every 
  $p(i) \in Q \times \Nset_0$ where $R(p(i)) = b$. This means
  that also $\calP(\Reach_{Q{\times}\{j\}}^\sigma(p(i))) = 1$ for every $j$
  such that $0 \leq j < i$. Now suppose that 
  $\calP(\Reach_{\White}^\sigma(p(i))) > 0$. Then there is some white
  configuration $q(j)$ such that $\calP(\Reach_{\{q(j)\}}^\sigma(p(i))) > 0$.
  Since $q(j)$ is white, we have that $\calP(\ST^\sigma(q(j))) < 1$.
  Thus, we obtain that $\calP(\ST^\sigma(p(i))) < 1$, which is a contradiction.
  Since $\calP(\Reach_{Q{\times}\{j\}}^\sigma(p(i))) = 1$ and
  $\calP(\Reach_{\White}^\sigma(p(i))) = 0$, we have that
  $\calP(\Reach_{\Black_{j}}^\sigma(p(i))) = 1$.
  \qed
\end{proof}

\begin{reftheorem}{Lemma}{\ref{lem-period}}
  There is $1 \leq \ell \leq N$ such that, for every 
  $j \geq N$, the columns $R_j = R_{j+\ell}$.
\end{reftheorem}
\begin{proof}
  We show that for all $j,k \in \Nset$ we have that if $R_j = R_k$,
  then also $R_{j+1} = R_{k+1}$. From this we easily obtain our 
  lemma---since there are at most $N$ different columns, there
  are $m,n \in \Nset$ such that $0 \leq m < n \leq N$ and
  $R_m = R_n$. We put $\ell = n - m$. Obviously, $R_j = R_{j+\ell}$  
  for every $j \geq m$. Since $m < N$, we are done.

  It suffices to prove that for every $i \in \Nset$, the column $R_{i+1}$ is
  completely determined by the column $R_i$ in the following sense:
  For every $q \in Q$ we have that $R_{i+1}(q) = b$ iff there is a 
  strategy $\sigma$ such that $\calP(\Reach_{\Black_i}^\sigma(q(i{+}1))) = 1$ and
  $\calP(\Reach_{\White_i}^\sigma(q(i{+}1))) = 0$. Note that the existence
  of $\sigma$ does not depend on the exact value of $i$ as long as
  the column $R_i$ stays the same. Hence, the above claim implies
  that if $R_j = R_k$, then also $R_{j+1} = R_{k+1}$. It remains to 
  prove this claim. The ``$\Rightarrow$'' direction
  follows directly from Lemma~\ref{lem-reach}.
  For the ``$\Leftarrow$'' direction,
  consider a strategy $\sigma$ such that 
  $\calP(\Reach_{\Black_i}^\sigma(q(i{+}1))) = 1$ and 
  $\calP(\Reach_{\White_i}^\sigma(q(i{+}1))) = 0$. 
  For each $p(i) \in \Black_i$ there is a strategy
  $\sigma_p$ such that $\calP(\ST^{\sigma_p}(p(i))) = 1$. Hence, we can
  construct a strategy $\pi$ which behaves like $\sigma$ until
  some $p(i) \in \Black_i$ is reached, and from that point on it behaves
  like $\sigma_p$. Obviously, $\calP(\ST^{\pi}(q(i{+}1))) = 1$ as needed.
\qed
\end{proof}

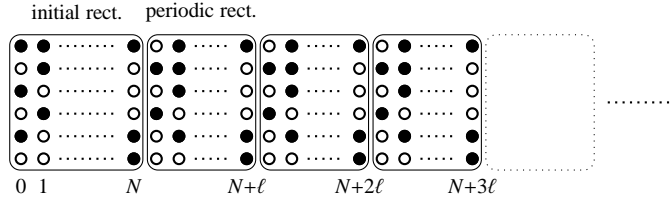
\begin{figure}[t]
\centering
\tikzstyle{w}=[circle,thick,draw,minimum size=1ex,inner sep=0em]
\tikzstyle{b}=[circle,thick,draw,fill=black,minimum size=1ex,inner sep=0em]
\begin{tikzpicture}[x=.3cm,y=.3cm,font=\scriptsize]
\foreach \q in {0,1,2,3,4,5}
  \foreach \i in {0,1,5,6,7,10,11,12,15,16,17,20} 
    {\node at (\i,\q) [w] {};}
\foreach \q in {0,1,5}
  \foreach \i in {5,10,15,20} 
    {\node at (\i,\q) [b] {};}
\foreach \q in {2,4}
  \foreach \i in {6,11,16} 
    {\node at (\i,\q) [b] {};}
\foreach \q in {1,3,4,5}
  \foreach \i in {7,12,17} 
    {\node at (\i,\q) [b] {};}
\node at (0,1) [b] {};
\node at (0,3) [b] {};
\node at (0,5) [b] {};
\node at (1,2) [b] {};
\node at (1,4) [b] {};
\node at (1,5) [b] {};
\foreach \q in {0,1,2,3,4,5}
   {\draw (1.7,\q) to [thick,dotted] (4.3,\q);
    \draw (7.7,\q) to [thick,dotted] (9.3,\q);
    \draw (12.7,\q) to [thick,dotted] (14.3,\q);
    \draw (17.7,\q) to [thick,dotted] (19.3,\q);}
\draw (-.5,-.5) [rounded corners] rectangle (5.4,5.5);
\draw (5.6,-.5) [rounded corners] rectangle (10.4,5.5);
\draw (10.6,-.5) [rounded corners] rectangle (15.4,5.5);
\draw (15.6,-.5) [rounded corners] rectangle (20.4,5.5);
\draw (20.6,-.5) [rounded corners,dotted] rectangle (25.4,5.5);
\node at (0,-1.2) {$0$};
\node at (1,-1.2) {$1$};
\node at (5,-1.2) {$N$};
\node at (10,-1.2) {$N{+}\ell$};
\node at (15,-1.2) {$N{+}2\ell$};
\node at (20,-1.2) {$N{+}3\ell$};
\node at (2.5,6.5) {initial rect.{}};
\node at (8,6.5) {periodic rect.{}};
\draw (26,2.5) to [thick,dotted] (29,2.5);
\end{tikzpicture}
\caption{The structure of coloring $R$ (where $N=2^{|Q|}$).}
\label{fig-R-coloring}
\end{figure}

Now we show that the initial and periodic rectangles of the coloring 
$R$ (given in Figure \ref{fig-R-coloring}) are computable
in exponential time. For this we need to formulate and prove an important
observation which establishes a powerful link to the results presented
in Section~\ref{sec-noboundary}. We start by defining 
a OC-MDP $\A_{R,\ell}$, which
encodes the structure obtained by deleting all white points from the 
periodic rectangle of $R$. Later, we construct such an automaton also
for another coloring $B$, where some points are gray. Therefore,
the definition of $\A_{R,\ell}$ is parametrized by a general coloring 
which satisfies certain conditions.

\begin{definition}[the OC-MDP $\A_{C,\ell}$]
\label{def-OC-AC}
Let $C : Q \times \Nset_0 \rightarrow \{b,w,g\}$ be a coloring such that
$C_{N} = C_{N+\ell}$ and for every $p(N{+}i) \in Q \times \Nset$ where
$1 \leq i \leq \ell$ and $C(p(N{+}i)) \neq w$ we have that
\begin{itemize}
\item[(1)] if $p(N{+}i)$ is probabilistic and $p(N{+}i) \ctran{} q(N{+}j)$, then
  $C(q(N{+}k)) \neq w$, where \mbox{$k = j \mod \ell$};
\item[(2)] if $p(N{+}i)$ is non-deterministic, then there is some 
  $p(N{+}i) \ctran{} q(N{+}j)$ such that
  $C(q(N{+}k)) \neq w$, where \mbox{$k = j \mod \ell$}.
\end{itemize}
We define a OC-MDP $\A_{C,\ell}$ where
\begin{itemize}
\item the set $Q_{C,\ell}$ of control states of $\A_{C,\ell}$ consists of all 
  $[p,i]$ where $p \in Q$, $1 \leq i \leq \ell$, and $C(p(N{+}i)) \neq w$.
  A given control state $[p,i]$ is non-deterministic or probabilistic,
  depending on whether $p \in Q_N$ or $p \in Q_P$, respectively;
\item the set of zero rules consists of all triples
  $([p,i],0,[p,i])$, where $[p,i] \in Q_{C,\ell}$;
\item the set of positive rules is constructed as follows:
  \begin{itemize}
  \item for all $(p,c,q) \in \delta^{>0}$ and all $i \in \Nset$ such 
    that $1 \leq i \leq \ell$, $1 \leq i{+}c \leq \ell$, and
    $[p,i],[q,i{+}c] \in Q_{C,\ell}$, we add a rule $([p,i],0,[q,i{+}c])$.
    If $[p,i]$ is probabilistic, then the probability of the rule 
    $([p,i],0,[q,i{+}c])$ is $P^{>0}(p,c,q)$. 
  \item for all $(p,c,q) \in \delta^{>0}$ and all $i \in \Nset$ such 
    that $1 \leq i \leq \ell$, $i{+}c = \ell{+}1$, and
    $[p,i],[q,1] \in Q_{C,\ell}$, we add a rule $([p,i],1,[q,1])$.
    If $[p,i]$ is probabilistic, then the probability of the rule 
    $([p,i],1,[q,1])$ is $P^{>0}(p,c,q)$. 
  \item for all $(p,c,q) \in \delta^{>0}$ and all $i \in \Nset$ such 
    that $1 \leq i \leq \ell$, $i{+}c = 0$, and
    $[p,i],[q,\ell] \in Q_{C,\ell}$, we add a rule $([p,i],-1,[q,\ell])$.
    If $[p,i]$ is probabilistic, then the probability of the rule 
    $([p,i],-1,[q,\ell])$ is $P^{>0}(p,c,q)$. 
  \end{itemize}
\end{itemize}
Observe that conditions~(1) and~(2) guarantee that $\A_{C,\ell}$
is indeed an OC-MDP.
\end{definition}

\begin{lemma}
\label{lem-CN-connection}
  For each configuration $[p,i](j)$ of $\A_{R,\ell}$ we have that
  \mbox{$\Val^{\NT}_{\Dz_{\A_{R,\ell}}}([p,i](j)) =1$}.
\end{lemma}
\begin{proof}
  Let $[p,i](j)$ be a configuration of $\A_{R,\ell}$. By definition
  of $\A_{R,\ell}$, we have that $R(p(N{+}i{+}j\ell)) = b$. 
  By Lemma~\ref{lem-reach}, there is a MD strategy $\sigma$ such
  that $\calP(\Reach_{\Black_{N}}^\sigma(p(i))) = 1$ and
  $\calP(\Reach_{\White}^\sigma(r(m))) = 0$ for every 
  $r(m) \in Q \times \Nset_0$ where $R(r(m))=b$. Consider a MD strategy
  $\pi$ in $\Dz_{\A_{R,\ell}}$ defined as follows: for every configuration
  $[q,k](n)$ of $\A_{R,\ell}$ where $q \in Q_N$ we put
  $\pi([q,k](n)) = [q',k'](n')$, where 
  \begin{itemize}
  \item $\sigma(q(N{+}k{+}n\ell)) = q'(t)$,
  \item $k' = (t-N) \mod \ell$,
  \item $n' = (t-N) \div \ell$.
  \end{itemize}
  Note that the definition of $\pi$ is correct, because $R(q'(t)) = b$
  and hence the transition $\pi([q,k](n)) = [q',k'](n')$ exists
  in $\Dz_{\A_{R,\ell}}$ (realize that if $R(q'(t))$ was white, we would
  have a contradiction with  
  $\calP(\Reach_{\White}^\sigma(q(N{+}k{+}n\ell))) = 0$). Since almost all
  runs of $\Dz_{\A}(\sigma)$ initiated in $p(N{+}i{+}j\ell)$ visit
  $\Black_N$, we obtain that almost all runs of 
  $\Dz_{\A_{R,\ell}}(\pi)$ initiated in $[p,i](j)$ visit a configuration
  of the form $[q,\ell](0)$. This means that 
  \mbox{$\Val^{\NT}_{\Dz_{\A_{R,\ell}}}([p,i](j)) =1$}.
  \qed
\end{proof}

\begin{lemma}
\label{lem-path-bound}
  Let $\A = (Q,\delta^{=0},\delta^{>0},(Q_N,Q_P),P^{=0},P^{>0})$ be a OC-MDP.
  If $p(i) \ctran{}^* q(0)$, then there is a path from $p(i)$ to $q(0)$
  in $\Dz_{\A}$
  such that the counter stays bounded by $i {+} |Q|^2$ along this path.
\end{lemma}
\begin{proof}
  For every $j \in \Nset_0$, we define a relation
  ${\leadsto_j} \subseteq Q \times Q$ inductively as follows:
  \begin{itemize}
  \item ${\leadsto_0} = \{(s,t) \in Q\times Q \mid s(1) \ctran{} t(0)\}$
  \item ${\leadsto_{j+1}}$ consists of all $(s,t) \in Q\times Q$ such that
    one of the following conditions is satisfied:
    \begin{itemize}
    \item $s \leadsto_j t$;
    \item $s(1) \ctran{} r(1)$ for some $r \in Q$ such that $r \leadsto_j t$;
    \item $s(1) \ctran{} r(2)$ for some $r \in Q$ such that 
       $r\leadsto_j u$ and $u \leadsto_j t$ for some $u \in Q$.
    \end{itemize}
  \end{itemize}
  A straightforward induction on $j$ reveals that if $s \leadsto_j t$,
  then there is a path from $s(1)$ to $t(0)$ in $\Dz_{\A}$ 
  along which the counter stays bounded by $j+1$.

  Let ${\leadsto} = \bigcup_{j \in \Nset_0} {\leadsto_j}$. Observe that
  ${\leadsto} = {\leadsto_{|Q|^2}}$. One can easily show that 
  $s \leadsto t$ iff for every $i \in \Nset$ there is a path from 
  $s(i)$ to $t(i{-}1)$ such that the counter is less or equal to 
  $i+|Q|^2$ and greater or equal to $i$ in all configurations except for the 
  last one (the ``$\Rightarrow$'' direction is proven for every ${\leadsto_j}$ 
  by induction on $j$, and the ``$\Leftarrow$'' direction is proven 
  by induction on the length of a path from $s(i)$ to $t(i{-}1)$). 
  From this we get that if there is a path from
  $s(i)$ to $t(0)$ in $\Dz_{\A}$ such that the counter stays positive in all
  configurations except for the last one, then there is a path from
  $s(i)$ to $t(0)$ along which the counter is bounded by $i+|Q|^2$.
  Finally, we show that if there is a path from
  $s(i)$ to $t(0)$ along which the counter becomes zero $m$ times,
  then there is a path from
  $s(i)$ to $t(0)$ along which the counter is bounded by $i+|Q|^2$
  (this is the result we are aiming at). However, this is easy to prove 
  by induction on $m$.
\qed
\end{proof}

\begin{lemma}
\label{lem-reg-strategy}
  There is a counter-regular strategy $\sigma$ which is optimal in 
  every configuration of $\OptValOne^{\ST}$. Further, the underlying
  $\A$-automaton and selector function of the strategy $\sigma$ 
  are computable from the initial and periodic rectangles of the
  coloring~$R$ in time which is exponential in the size of $\A$. 
\end{lemma}
\begin{proof}
  We design a MD strategy $\pi$ such that 
  \begin{itemize}
    \item $\pi$ is optimal in every configuration of $\OptValOne^{\ST}$.
    \item $\pi(p(i)) = \pi(p(i{+}\ell))$ for all $p \in Q_N$ and 
       $i > |Q|^2N^2 + N$.
    \item $\pi(p(i))$ is computable for all $p \in Q_N$
       and $i \leq |Q|^2N^2 + N + \ell$ in time polynomial in~$N$, assuming
       that the initial and periodic rectangles of $R$ are known. 
  \end{itemize}
  Obviously, the strategy $\pi$ can be easily transformed into
  a counter-regular strategy $\sigma$ with the required properties.

  First, for every $p(i)$ such that $i \leq N$ and $R(p(i)) = b$ we fix a 
  finite path $p(i) \ctran{} \cdots \ctran{} q(0)$
  where $q \in F$ and all configurations in the path are black in $R$.
  Such a path must exist, and we can further safely assume that 
  the counter stays bounded by $|Q|^2N^2 + N$ along this path 
  (see Lemma~\ref{lem-path-bound}) and no
  configuration appears twice in the path. 
  For all configurations $q(0)$ where $q \in F \cap Q_N$, the strategy
  $\pi$ is defined arbitrarily.
  Now, for each path $w$ fixed above
  (in any order) we do the following: we identify all non-deterministic
  configurations $q(j)$ in $w$ for which the strategy $\pi$ has not yet
  been defined, and let $\pi(q(j))$ to select the (only) outgoing 
  transition of $q(j)$ that appears in the path $w$. Let $\pconf$ be
  the set of all configurations (non-deterministic or probabilistic)
  that appear in some of the finite paths fixed above. 

  Now consider again the OC-MDP $\A_{R,\ell}$. According to
  Theorem~\ref{thm-NT} and Lemma~\ref{lem-CN-connection}, there is a
  $\SMD$ strategy $\xi$ in $\Dz_{\A_{R,\ell}}$ such that for every
  configuration $[p,i](j)$ of $\A_{R,\ell}$ we have that 
  $\calP(\NT^{\xi}([p,i](j))) = 1$. For every control state $[p,i]$
  of $\A_{R,\ell}$ where $p \in Q_N$, let 
  $[p,i](1) \ctran{} [q,j](k)$ be the transition selected by
  $\xi([p,i](1))$. For every  $p(N{+}i{+}y\ell)$ such that
  $y \in \Nset_0$ and 
  $\pi(p(N{+}i{+}y\ell))$ has not yet been defined, we let   
  $\pi(p(N{+}i{+}y\ell))$ to select the transition
  $p(N{+}i{+}y\ell) \ctran{} q(N{+}j{+}y\ell{+}(k{-}1)\ell)$.  

  Obviously, we have that $\pi(p(i)) = \pi(p(i{+}\ell))$ for all 
  $p \in Q_N$ and $i > |Q|^2N^2 + N$. If the initial and periodic rectangles
  of~$R$ are known, the automaton $\A_{R,\ell}$ is effectively
  constructible by using Definition~\ref{def-OC-AC}, and the
  $\SMD$ strategy $\xi$ is computable in time polynomial in~$N$
  by Theorem~\ref{thm-NT}. Hence, $\pi(p(i))$ is computable 
  for all $p \in Q_N$ and $i \leq |Q|^2N^2 + N + \ell$ in time polynomial 
  in~$N$. To see that $\pi$ is optimal in every configuration of
  $\OptValOne^{\ST}$, realize the following:
  \begin{itemize}
    \item Let $\White = \{q(j) \in Q \times \Nset_0 \mid R(q(j))= w\}$.
       Then for every $p(i) \in \OptValOne^{\ST}$ we have that
       $\calP(\Reach_{\White}^{\pi}(p(i))) = 0$. 
    \item Let $\fin = \{q(0) \mid q \in F\}$. Then there is a fixed 
       $\varepsilon > 0$ such that for every $p(i) \in \pconf$ we have
       that $\calP(\Reach_{\fin}^{\pi}(p(i))) \geq \varepsilon$. This
       is because for each of the finitely many $p(i) \in \pconf$
       there is a finite path from $p(i)$ to $\fin$ in $\Dz_{\A}(\pi)$.
    \item For each $p(i) \in \OptValOne^{\ST} \setminus \pconf$
       we have that \mbox{$\calP(\Reach_{\pconf}^{\pi}(p(i))) = 1$}. This is 
       because almost all runs in $\Dz_{\A}(\pi)$ initiated in $p(i)$ tend to 
       decrease the counter until they reach a configuration of $\pconf$. 
  \end{itemize}
  From these three properties, one can conclude that 
  $\calP(\ST^{\pi}(p(i))) = 1$ for every $p(i) \in \OptValOne^{\ST}$.
\qed
\end{proof}

\begin{figure*}[t]
{
\begin{tabbing}
\hspace*{2em} \=  \hspace*{1.2em} \= \hspace*{1.2em} \= \hspace*{1.2em} 
     \= \hspace*{1.2em} \= \hspace*{1.2em} \= \hspace*{1.2em} \= \kill
\makebox[4em][l]{\textbf{Input:}} An OC-MDP 
$\A =(Q,\delta^{=0},\delta^{>0},(Q_N,Q_P),P^{=0},P^{>0})$, a non-empty set
$F \subseteq Q$ of final states.\\
\makebox[4em][l]{\textbf{Output:}} The initial and periodic 
rectangles of the coloring $R$.\\[1ex]
\texttt{1:} \> \FOREACH $p(i)$ where $0 \leq i \leq 2N$ \DO 
    $A(p(i)) := w$ \DONE\\
\texttt{2:} \> \FOREACH $\ell$ where $1 \leq \ell \leq N$ \DO\\ 
\texttt{3:} \>\> \FOREACH $C$ where $C : Q \rightarrow \{b,w\}$ \DO\\
\texttt{4:} \>\>\> \FOREACH  $p(i)$ where $0 \leq i \leq N+\ell$ \DO
    $B(p(i)) := g$ \DONE\\
\texttt{5:} \>\>\> \FOREACH  $q \in F$ \DO $B(q(0)) := b$ \DONE\\
\texttt{6:} \>\>\> $B_N := C;$\ \  $B_{N+\ell} := C$\\
\texttt{7:} \>\>\> \REPEAT\\
\texttt{8:} \>\>\>\> \FOREACH $p(i)$ where $0 \leq i \leq N+\ell$ \DO
    $B(p(i)) := \cc(p(i))$ \DONE\\
\texttt{9:} \>\>\> \UNTIL $B$ does not change\\
\texttt{10:}\>\>\> \IF $B(p(i)) = r$ for some $p(i)$ \THEN \CONTINUE
    with the next $C$\\
\texttt{11:}\>\>\> compute the OC-MDP $\A_{B,\ell}$\\
\texttt{12:}\>\>\> \FOREACH $p(N{+}i)$ where $1 \leq i \leq \ell$ and
                   $B(p(N{+}i)) \neq w$ \DO\\
\texttt{13:}\>\>\>\> $B(p(N{+}i)) = \cv(p(N{+}i))$\\
\texttt{14:}\>\>\> \DONE\\
\texttt{15:}\>\>\> \IF $B(p(i)) = r$ for some $p(i)$ \THEN \CONTINUE 
                         with the next $C$\\
\texttt{16:}\>\>\> \REPEAT\\
\texttt{17:}\>\>\>\> \FOREACH $p(i)$ where $0 \leq i \leq N$ \DO
    $B(p(i)) := \cp(p(i))$ \DONE\\
\texttt{18:}\>\>\> \UNTIL $B$ does not change\\
\texttt{19:}\>\>\> \FOREACH $p(i)$ where $0 \leq i \leq N$ and 
                        $B(p(i)) = g$ \DO $B(p(i)) := b$ \DONE \\
\texttt{20:}\>\>\> \IF $B(p(i)) = r$ for some $p(i)$\\
\texttt{21:}\>\>\>\> \THEN \CONTINUE with the next $C$\\
\texttt{22:}\>\>\>\> \ELSE transfer all black points of $B$ to $A$\\
\texttt{23:}\>\> \DONE\\
\texttt{24:}\> \DONE\\
\texttt{25:}\> find the least $\ell$ such that $A_N = A_{N{+}\ell}$\\
\texttt{26:}\> \OUTPUT $A_0,\ldots,A_N$ and $A_{N{+}1},\ldots,A_{N{+}\ell}$
\end{tabbing}
}
\caption{An exponential-time algorithm which computes the coloring $R$} 
\label{fig-alg-R}
\end{figure*}

\begin{reftheorem}{Theorem}{\ref{thm-optvalone}}
  An $\A$-automaton recognizing the set $\OptValOne^{\ST}$ is computable
  in exponential time. Further, there is a counter-regular strategy $\sigma$
  constructible in exponential time which is optimal in every
  configuration of $\OptValOne^{\ST}$.
\end{reftheorem}
\begin{proof}
  To construct an $\A$-automaton recognizing the set $\OptValOne^{\ST}$,
  it suffices to compute the initial and periodic rectangles of~$R$.
  This is achieved by the algorithm given in Fig.~\ref{fig-alg-R}.
  
  Since the width of the initial rectangle
  is $N+1$ and the width of the periodic rectangle is at most $N$,
  it suffices to compute the first $2N + 1$ columns of $R$. For 
  this purpose, we introduce two auxiliary colorings $A$ and $B$ whose domain
  is restricted to \mbox{$Q \times \{0,\ldots,2N\}$}. The coloring
  $A$ is just a memory used to accumulate the information about all of the newly 
  discovered black points. The color of all points in $A$ 
  is initially white (line~1) and, as we shall see, each
  $p(i)$ such that $0 \leq i \leq 2N$ and $R(p(i)) = b$ is eventually
  recolored to black in~$A$ at line~22. 

  The coloring $B$ is used to
  discover more and more points that are black in $R$. This is achieved
  by trying out all candidates $\ell$ for the width of the periodic
  rectangle (line~2) and all candidates $C$ for the column 
  $R_{N{+}\ell}$ (line~3). For each choice of $\ell$ and $C$, the 
  color of all $p(i)$ in $B$, where $0 \leq i \leq N{+}\ell$,
  is first initialized to gray at line~4 (the intuitive meaning of gray 
  is ``don't know''). Then, all $q(0)$ where $q \in F$ are recolored
  to black at line~5, which is surely correct. Further, the columns
  $B_{N{+}\ell}$ and $B_{N}$ are set to the current candidate $C$ (note
  $R_{N{+}\ell} = R_N$). Now, we try to recolor as much points as we can
  using the function $\cc$ (lines~7--9). For a given $p(i)$, 
  where $0 \leq i \leq N{+}\ell$, the function $\cc$ first computes
  the set of $\col(p(i))$ of colors that $p(i)$ should have according
  to its $\ctran{}$ successors and predecessors 
  (we say that $q(j)$ is a $\ctran{}$ successor of $r(k)$ if 
  $r(k) \ctran{} q(j)$). Formally, $\col(p(i))$ is the least set of
  colors satisfying the following:
  \begin{itemize}
  \item if $p \in Q_P$ and all $\ctran{}$ successors of $p(i)$
    are black in $B$, then $b \in \col(p(i))$;
  \item if $p \in Q_P$ and some $\ctran{}$ successor of $p(i)$
    is white in $B$, then $w \in \col(p(i))$;
  \item if $p \in Q_N$ and all $\ctran{}$ successors of $p(i)$
    are white in $B$, then $w \in \col(p(i))$;
  \item if $p \in Q_N$ and some $\ctran{}$ successor of $p(i)$
    is black in $B$, then $b \in \col(p(i))$;
  \item if $q(j) \ctran{} p(i)$ where $q \in Q_P$ and $q(j)$ is black in $B$,
    then $b \in \col(p(i))$;
  \item if $q(j) \ctran{} p(i)$ where $q \in Q_N$ and $q(j)$ is white in $B$,
    then $w \in \col(p(i))$.
  \end{itemize}
  Note that in the case when $i = N{+}\ell$, we need to know the $B$ color
  of $\ctran{}$ successors and predecessors of $p(i)$ whose counter
  value can also be $N+\ell+1$. Here we stipulate that 
  $B(q(N{+}\ell{+}1)) = B(q(N{+}1))$ (note that 
  $R(q(N{+}\ell{+}1)) = R(q(N{+}1)$). Intuitively, $\cc(p(i))$ contains 
  the color of $p(i)$ that is ``enforced'' by the colors of its 
  $\ctran{}$ successors and predecessors. If both black and white is 
  enforced, or if $B(p(i))$ is inconsistent with the enforced color,
  we discovered an inconsistency in the current choice of $\ell$ and $C$.
  Hence, the color which is returned by $\cc(p(i))$ is determined as follows:
  \begin{itemize}
  \item if $\col(p(i)) = \emptyset$, then $\cc(p(i))$ returns 
     $B(p(i))$ (i.e., the current color of $p(i)$ in $B$);
  \item if $\col(p(i)) = \{c\}$ and $B(p(i))=g$, then $\cc(p(i))$ returns~$c$;
  \item if $\col(p(i)) = \{c\}$ and $B(p(i))=c$, then $\cc(p(i))$ returns~$c$;
  \item in the other cases, $\cc(p(i))$ returns~$r$.
  \end{itemize}
  Note that the red color is used to mark a consistency error.
  Also note that each $p(i)$ is recolored at most twice, and so the
  \textbf{repeat-until} loop in lines~7--9 terminates after 
  $\calO(N)$ iterations, where each iteration invokes the function
  $\cc$ only $\calO(N)$ times.

  After terminating the loop in lines~7--9, the algorithm checks if 
  there is a red $p(i)$ and if it is the case, it rejects the current
  $C$ and continues with the next candidate (line~10). Otherwise,
  all points in $B$ are either black, white, or gray, where
  \begin{itemize}
  \item[(1)] for all $p(i)$ such that $B(p(i)) = g$ we have that 
    $\cc(p(i))$ returns $g$;
  \item[(2)] for all $p(i)$ such that $B(p(i)) \neq g$ we have that 
    if the width of the periodic rectangle of $R$ is $\ell$ and
    $R_{N{+}\ell} = C$ (i.e, the current candidates $\ell$ and $C$ are 
    the ``real'' ones), then $B(p(i)) = R(p(i))$. It is easy to show 
    that this claim is an invariant of the \textbf{repeat-until} 
    loop in lines~7--9.
  \end{itemize}
  Now we need to resolve the color of the remaining gray points.
  First, we concentrate on the gray points in the columns
  $B_{N{+}1},\ldots,B_{N{+}\ell}$ and check whether they can constitute
  the periodic rectangle of $R$ after some further recoloring.
  This is done by checking the condition of Lemma~\ref{lem-CN-connection}.
  First we construct the OC-MDP $\A_{B,\ell}$ of 
  Definition~\ref{def-OC-AC} (line~11). Note that the condition~(2)
  above guarantees that the coloring~$B$
  satisfies the requirements of Definition~\ref{def-OC-AC}.
  For each $p(N{+}i)$ where
  $1 \leq i \leq \ell$ and $B(p(N{+}i)) \neq w$ we 
  recolor $p(N{+}i)$ to $\cv(B(p(N{+}i)))$ at lines~12--14. Here 
  the function $\cv$ does the following: if $B(p(N{+}i)) = g$, 
  then $\cv(B(p(N{+}i)))$ returns either $b$ or $w$ depending on whether 
  $\Val^{\NT}_{\Dz(\A_{B,\ell})}([p,i](j)) = 1$ for all $j \in \Nset_0$ or not, 
  respectively. If $B(p(N{+}i)) \neq g$, then $\cv(B(p(N{+}i)))$ returns either
  $b$ or $r$, depending on whether 
  $\Val^{\NT}_{\Dz(\A_{B,\ell})}([p,i](j)) = 1$ for all $j \in \Nset_0$
  or not, respectively. Note that $\cv(B(p(N{+}i)))$ is computable 
  in time polynomial in the size of $N$ by Theorem~\ref{thm-NT}.
  Then we check whether some point has been
  recolored to red, and if it is the case, we continue with the next
  candidate (line 15). Otherwise, all points in the columns
  $B_{N{+}1},\ldots,B_{N{+}\ell}$ are now black or white. It is important
  to note that the functions $\cc$ and $\cv$ would not report any 
  inconsistencies in the current $B$ (i.e., if we run the code at 
  lines~7--14 again after line~15, no point would be recolored to red).
  This follows directly from Remark~\ref{rem-NT-structure}.
  
  It remains to resolve the gray points in the columns 
  $B_0,\ldots,B_N$. Here we use the observation about $R$ formulated
  in Lemma~\ref{lem-reg-strategy}.  Let $\hat{B}$ be the (only) 
  coloring satisfying the following conditions:
  \begin{itemize} 
  \item $\hat{B}_j = B_j$ for every  $0 \leq j \leq N{+}\ell$;
  \item $\hat{B}_{N{+}\ell{+}i} = \hat{B}_{N{+}i}$ for every $i \in \Nset$. 
  \end{itemize}
  For every  $p(i)$ where $0 \leq i \leq N$ and $B(p(i)) \neq w$, we 
  recolor $p(i)$ to $\cp(p(i))$. The function $\cp(p(i))$ checks,
  depending on whether $p$ is probabilistic/non-deterministic, whether 
  for all/some $p(i) \ctran{} r(j)$ there
  is a finite path $r(j) \ctran{} \cdots \ctran{} q(0)$  
  such that $q \in F$ and all configurations in this path are black or gray in
  the current $\hat{B}$. If this is the case, $\cp(p(i))$ returns
  the current $B(p(i))$. Otherwise, $\cp(p(i))$ returns either
  white or red, depending on whether $B(p(i)) = g$ or $B(p(i)) = b$,
  respectively. After finishing the loop at lines~16--18, 
  all of the remaining gray points of $B_0,\ldots,B_N$ are recolored to
  black at line~19.   
  Note that the function $\cp$ can be implemented in time polynomial
  in $N$ by employing, e.g., standard polynomial-time algorithms 
  for the reachability problem in pushdown automata.  
  Then we check whether some point has been
  recolored to red, and if it is the case, we continue with the next
  candidate (line 21). Otherwise, all points of $B_0,\ldots,B_{N{+}\ell}$  
  are black or white. Observe that
  \begin{itemize}
  \item for every $p(i)$ such that $i \leq N$ and $B(p(i)) = b$ 
    there is a finite path $p(i) \ctran{} \cdots \ctran{} q(0)$
    where $q \in F$ and all configurations in the path are black in 
    $\hat{B}$. Further, if $p \in Q_P$ and $p(i) \ctran{} r(j)$, then
    $B(r(j)) = b$.
  \item there is a $\SMD$ strategy $\xi$ in $\Dz_{\A_{B,\ell}}$ such 
    that for every configuration $[p,i](j)$ of $\A_{R,\ell}$ we have that 
    $\calP(\NT^{\xi}([p,i](j))) = 1$.
  \end{itemize}
  These are \emph{exactly} the ingredients which were needed to construct
  the strategy $\pi$ in the proof of Lemma~\ref{lem-reg-strategy}.
  If we apply the same construction to the coloring $\hat{B}$, we
  obtain a strategy $\pi_B$ such that $\calP(\ST^{\pi_B}(p(i))) = 1$ 
  for every $p(i) \in Q \times \Nset_0$ where $\hat{B}(p(i)) = b$.
  This means that all black points in the columns $B_0,\ldots,B_{N{+}\ell}$
  can be safely transferred from $B$ to $A$, which is done at line~22.

  After terminating the loop at lines 2--24, the algorithm finds the
  least $\ell$ such that $A_{N} = A_{N+\ell}$, and outputs the rectangles
  $A_0,\ldots,A_N$ and $A_{N{+}1},\ldots,A_{N{+}\ell}$. Since the ``real''
  values of $\ell$ and $C$ are eventually tested as candidates and the
  algorithms recolors a gray point to a white point only if some
  condition satisfied by $R$ is violated, all black points of 
  $R_0,\ldots,R_{N{+}\ell}$ are eventually discovered. 
  Since the functions $\cc$, $\cv$, and $\cp$ need only polynomial
  time in the size of $N$, the whole algorithm is polynomial in the
  size of $N$.

  After computing the initial and periodic rectangles of $R$,
  a counter-regular strategy $\sigma$ which is optimal for all
  configurations of $\OptValOne^{\ST}$ can be constructed 
  by using Lemma~\ref{lem-reg-strategy}.
\end{proof}

\begin{reftheorem}{Theorem}{\ref{thm-term-hard}}
  Membership in $\ValOne^{\ST}$ is \textbf{BH}-hard.
  Membership in $\OptValOne^{\ST}$ is $\PSPACE$-hard.
\end{reftheorem}

\begin{proof}\mbox{}
We start with proving the \textbf{BH}-hardness.
Our proof is essentially a variation on a proof by Serre
\cite{Serre06} (using a technique that 
originated in \cite{Kuc03} and was later reshaped in \cite{JKMS04})
showing that the reachability problem for 
non-probabilistic 2-player 1-counter games is
\DP-hard.
We show that similar arguments work to show \BH-hardness for OC-MDPs.

First, we show that membership in
$\ValOne^{ST}$
is
\NP-hard and \coNP-hard, 
and then we show how to combine these to get \BH-hardness.

We start with \NP-hardness. We reduce from SAT.  Suppose we are given
a CNF formula
$\psi = C_1 \wedge \ldots \wedge C_m$,
over variables $\{x_1,\ldots,x_r\}$.
We will encode assignments to the variables of $\psi$ by 
integers, as follows. Let $\pi_1,\ldots,\pi_r$ denote the first $n$ prime
numbers. Then an integer $n$ corresponds to an assignment that assigns true to $x_i$ if
and only if $\pi_i$ divides $n$.
Note that multiple integers map to the same assignment, but that all
assignments are certainly mapped to by some positive integer
(e.g., $1$
assigns false every variable).  It follows from the strong forms of
Bertrand's postulate (see, e.g., Theorem~5.8 in \cite{Shoup08}) that
(as a very conservative bound), for all $r \geq 64$, $\pi_r \leq
(2r)^2$.  (We can thus of course trivially compute the first $r$
primes $\pi_1, \ldots, \pi_r$ in time polynomial in $r$.)

The OC-MDP will have a start
state $s_0$, which is
controlled by the (maximizing)
player.   The  initial configuration is  $s_0(1)$ and the player
can choose to increment the counter and stay in state
$s_0$, or to move to state $s_1$ without changing the counter.  Thus,
after it has repeatedly incremented the counter up to a ``guessed''
number $n \geq 0$ which represents an assignment, the
game moves to configuration $s_1(n)$. 

State $s_1$ is probabilistic,
and it chooses,
uniformly at random, one of the clauses $C_i$, which it claims is 
not satisfied by the assignment associated with $n$, and
moves
to  configuration $s'_i(n)$.  $s'_i$ is controlled by the maximizing player,
and it
chooses a literal $l_j$ in $C_i$, and moves to 
$s'_{i,l_j}(n)$.
Suppose $l_j = x_j$.
From this configuration we 
deterministically
decrement the counter, but keep track, using $\pi_j$ auxiliary
states, how  many times, $\bmod{\ \pi_j}$, we have decremented the counter.
Clearly, if we hit the counter value $0$  in a state that
indicates we have decremented a number of times which is $0 \pmod{\pi_j}$,
then the assignment corresponding to $n$ satisfies clause $C_i$.
Similarly, if $l_j = \neg x_j$, we can check that the number of
times decremented is $\neq 0 \pmod{\pi_j}$, in which case again $n$ satisfies clause $C_i$.
Since the random
player
chose all clauses with equal probability,
there is
a strategy to terminate in such ``accepting'' 
states with probability~$1$ if
there is a 
satisfying assignment to $\psi$.
Also note that if  there is no
satisfying assignment to $\psi$, then
there is a fixed $\delta > 0$
such that for every strategy
the probability of non-terminating
or terminating in a ``non-accepting'' control state is at least~$\delta$.
Note that, as it is easy to check using the bound $\pi_r \leq (2r)^2$, 
the size of the resulting 1C-MDP is polynomial in the
size of the formula $\psi$.

Next, for \coNP-hardness, suppose we have a CNF formula
$\psi = C_1 \wedge \ldots \wedge C_m$,
over variables $\{x_1,\ldots,x_r\}$, and we
want to decide unsatisfiability.
We do as before, but with some role
reversals between non-deterministic and probabilistic control states.
Starting in configuration
$s_0(1)$ where $s_0$ is now probabilistic, 
we randomly either
increment the counter or
change the state to
$s_1$ 
(with, say, equal probability).  Thus we eventually move to state $s_1$ with
probability~$1$, and for every positive integer~$n$, with some positive
probability we move to
$(s_1,n)$. The state $s_1$ is controlled (i.e., non-deterministic).

The player's strategy
chooses (guesses) a clause $C_i$ which
it thinks
cannot be satisfied by the assignment $n$, and moves to configuration
$s'_i(n)$, where $s'_i$ is probabilistic.
Then
the random player picks
one of the literals $l_j$, of clause
$C_i$,
uniformly at random
(intuitively claiming 
at least one of
them
will be satisfied and thus with
positive probability we will terminate in a rejecting state),
and moves to 
$s'_{i,l_j}(n)$.
We then decrement deterministically as before, except that now when we terminate we accept
precisely in those states where we would have not accepted before.
Specifically, we accept if ``assignment'' $n$ did not assign true to
literal $l_j$ of clause $C_i$, which again we can check by keeping
track of how many times we decremented $\bmod{\ \pi_i}$, upon hitting
counter value~$0$.

Note that under every strategy the probability of termination is $1$.
Similarly as before,
there is a strategy
such that  the probability of termination in an accepting state is $1$
if there is no satisfying assignment to $\psi$,
on the other hand there is some $\delta>0$ such that
terminating in a ``non-accepting'' state occurs
with probability at least $\delta$ under every strategy
if there is a satisfying assignment to $\psi$.

Finally, to show \BH-hardness, consider any statement which is a
$\wedge$-$\vee$ combination
of statements of the form ``$\psi_i$ is satisfiable'' and
``$\psi_j$ is not-satisfiable'',
where $\psi_i$'s are Boolean formulas.
Deciding whether such statements are true is \BH-complete.
In order to mimic this with a OC-MDP, we do as follows:  $\vee$ is mimicked by
the controller (i.e., a non-deterministic state)
picking one of the disjuncts.
$\wedge$ is mimicked by
the random player (a probabilistic state)
picking one of the conjuncts uniformly at random.  
When we hit a statement ``$\psi_i$ is (un)satisfiable'', we play the
corresponding game.
It is easy to check that
maximizer has
a strategy to terminate in an accepting 
state with probability~$1$ if the entire statement is true,
and that there is a $\delta>0$ such that for every strategy
termination in an accepting state has probability at most $1-\delta$
if the entire statement is false.

\smallskip

Note that in all the OC-MDP from the reductions above the sets
$\OptValOne^{\ST}$ and $\ValOne^{\ST}$ are equal.
Thus we have already proved also \textbf{BH}-hardness of
the membership in both of them.
We will now prove, however, that the membership in
$\OptValOne^{\ST}$ is even $\PSPACE$-hard.

The proof is by reduction from the emptiness problem for
simple alternating finite automata over a one-letter alphabet.
A simple alternating finite automaton over a one-letter alphabet
(call it AFA for short in the rest of the text)
is a tuple
$(Q,\delta,q_0,F)$ where $Q$ is a finite nonempty set of \emph{states},
$q_0\in Q$, $F\subseteq Q$ and $\delta$ is a \emph{transition function}
assigning to every state either another state, or the ``existential''
pair $p\lor q$ of states $p,q\in Q$, or the ``universal'' pair
$p \land q$.
The automaton is used to recognise sets of words over a one-letter alphabet.
Such words can be considered as numbers from $\Nset_0$.
The language of the automaton is defined to be the set of exactly
those $n\in\Nset_0$ which are accepted from the state $q_0$,
written $Acc(q_0,n)$. The semantics of the expression $Acc(q,n)$,
meaning accepting a number $n$ from
a state $q$, is defined inductively on $n$:
$Acc(q,0)$ is true iff $q\in F$.
For $n=k+1$ we have three cases:
\begin{itemize}

\item If $\delta(q)=p$ then $Acc(q,k+1)$ is equivalent to $Acc(p,k)$.

\item If $\delta(q)=p_1\lor p_2$ then $Acc(q,k+1)$ is true iff at least one of
$Acc(p_1,k)$ and
$Acc(p_2,k)$ is true.

\item If $\delta(q)=p_1\land p_2$ then $Acc(q,k+1)$ is true iff both
$Acc(p_1,k)$ and
$Acc(p_2,k)$ are true.

\end{itemize}
See~\cite{JS07} for more details about AFA.
Proposition~4 from~\cite{JS07} states that
the problem of deciding whether the language of a given
AFA is empty, is
$\PSPACE$-hard.

We now describe a log-space reduction of the emptiness problem for AFA
to the membership in $\OptValOne^\ST$ for OC-MDP.
Let $(Q,\delta,q_0,F)$ be an AFA.
The reduction returns the following OC-MDP:
$(Q\cup\{p\},\delta^{=0},\delta^{>0},(Q_N,Q_P),P^{=0},P^{>0})$
along with the set $F$ of final states and the initial configuration
$p(1)$
where
\begin{itemize}

\item $p$ is a fresh new state, $p\notin Q$;

\item $\delta^{=0}
=\{(p,0,p)\}
\cup\{(q,0,q) \mid q\in Q\}$;

\item $\delta^{>0}
=\{(p,+1,p),(p,-1,q_0)\}
\cup\{(q,-1,r) \mid q,r\in Q,\text{whenever }r\text{ occurs in }\delta(q)\}$;

\item $Q_N=\{p\}\cup\{q\in Q \mid \exists r,s\in Q: \delta(q)=r\lor s\}$, $Q_P=Q\smallsetminus Q_N$;

\item the probability assignments always return the uniform distribution.

\end{itemize}
If $n$ is accepted by the AFA then the following MD
strategy $\sigma$ proves $p(1)\in\OptValOne^\ST$:
\begin{itemize}

\item $\sigma(p(n+1))=q_0(n)$ and
$\sigma(p(k))=p(k+1)$ for $k\neq n+1$,

\item
$\sigma(q(k))=r(k-1)$
for every $q\in Q_N\cap Q$ and $k\in\Nset$
where $r$ is an arbitrary state occurring in $\delta(q)$
with $Acc(r,k-1)$ being true, and

\item
$\sigma(q(k))$ is defined
arbitrarily if there is no such $r$.

\end{itemize}
On the other hand, if $\sigma$ ensures almost sure
reaching $F\times\{0\}$ from $p(1)$, there must be some
$n$ such that $q_0(n)$ is visited on some path from
$p(1)$ to $F\times\{0\}$ with positive probability.
It can easily be shown that every configuration
of the form $q(k)$ visited after $q_0(n)$ satisfies
$Acc(q,k)$. In particular $Acc(q_0,n)$ and thus the language
of the AFA is not empty.
\qed
\end{proof}

We now show that {\em qualitative} problems for the special 
subclass of OC-MDPs given by  {\em solvency games} \cite{BKSV08}
can be solved in polynomial time.
We now recall more formally 
the definition of solvency games
from \cite{BKSV08}, which was
described informally in the introduction.
A {\em solvency game}, is given by a positive integer, $n$, 
($n$ is the initial pot of money belonging to the gambler), and
a finite 
set $\A = \{A_1, \ldots, A_k\}$ of {\em actions} (or ``gambles''),
each of which is associated with 
a finite-support probability distribution on
the integers.  Since
for computational purposes we 
have to be given these distributions as finite input,
we
assume that the distribution associated with 
each action $A_i$, $i=1,\ldots,k$, 
is encoded by giving a set of pairs 
$\{ (n_{i,1},p_{i,1}), (n_{i,2},p_{i,2}), \ldots,
(n_{i,m_i}, p_{i,m_i})\}$,  such that 
for $j = 1,\ldots, m_i$,
$n_{i,j} \in \Zset$  and $p_{i,j}$ are positive rational probabilities,  
i.e., $p_{i,j} \in (0,1]$ 
and $\sum^{m_i}_{j=1} p_{i,j} = 1$.
We assume the integers $n_{i,j}$ and the rational
values $p_{i,j}$ are both encoded in the standard way, 
in {\em binary} notation.

In a solvency game the player (or {\em gambler} or {\em investor})
starts with the initial pot of money, $n$, and has to repeatedly choose
an action (gamble) from the set $\A$.
If at any time the current pot of money is $n'$, and the gambler 
then chooses action
$A_i$, then we sample from the finite-support 
distribution associated with $A_i$, and the integer, $d$, resulting from this
random sample is added
to $n'$, obtaining the new pot of money $n'+d$.  
If the pot of money hits $0$ or goes below zero,
then the gambler loses (goes bankrupt) and the game ends.
Otherwise, we repeat the gambling process with the new pot of money $n'+d$.
The gambler's aim is to minimize the probability of ever losing the game,
i.e., to  minimize the probability of ever going bankrupt.
(Note that we do not allow the gambler to simply choose to stop gambling
 (which would be too easy a way to prevent going bankrupt). 
 Our gamblers are hopelessly addicted!  Perhaps then {\em investor} is more appropriate.)

It should be clear that solvency games constitute a special subclass
of
OC-MDPs.
Namely,
the counter in an OC-MDP can be used to keep track of the gambler's wealth.
Although, by definition, OC-MDPs can only increment or decrement
the counter by one in each state transition, it is easy to 
augment any finite change to the counter value by using additional
states and incrementing or decrementing the counter by one at a time.  
Namely, the OC-MDP will have a ``base'' {\em control} state, $s$, from which
is chooses from the set of actions $\{A_1,\ldots,A_k\}$.
If action $A_i$, is associated with a probability
distribution given by 
$\{ (n_{i,1},p_{i,1}), (n_{i,2},p_{i,2}), \ldots,
(n_{i,m_i}, p_{i,m_i})\}$,  we will have $|n_{i,j}|$ additional auxiliary
states associated with each such integer $n_{i,j}$ in the support of $A_i$.
After the gambler chooses action $A_i$, we transition from state $s$ to
a new {\em random} state $s_i$ without changing the counter value. From
$s_i$ we move with probability $p_{i,j}$ to a new state $s_{i,j}$,
from which we will deterministically (with probability 1) 
add $n_{i,j}$ to the counter, 
doing the incrementing or decrementing one at a time, by going
through $n_{i,j}$ additional states $s_{i,j,1}$, \ldots, $s_{i,j,n_{i,j}}$.
Finally, after this is done we return to the ``base'' control state $s$.
It is easy to see that the original solvency game with the 
objective of minimizing the probability of bankruptcy is equivalent to the
resulting
OC-MDP, started in state $s$, with the objective of minimizing the 
probability of ever reaching counter value $0$ (in {\em any} state).
Note that since we assume the integers $n_{i,j}$ are encoded in binary, 
in principle this reduction yields an OC-MDP that is 
exponentially larger than the input solvency game.  
Of course, to make this a polynomial time
reduction we can simply 
assume that the integers $n_{i,j}$ are encoded in unary.
Nevertheless, we show that even when the $n_{i,j}$'s are encoded in
binary, all qualitative problems for solvency games are decidable
in polynomial time: 

\begin{reftheorem}{Proposition}{\ref{prop-solvency}}
Given a solvency game, it is decidable in polynomial time
whether the gambler has a strategy to go bankrupt with  
probability: \  $> 0$, \ $=1$,\  $=0$, \ or \ $<1$.
\end{reftheorem}

\begin{proof}
The first three cases ( $> 0$, $=1$, $=0$) are either
trivial, or follow fairly easily from
what we have established about OC-MDPs, so we do these first.
The last case, $< 1$, is not easy at all, but follows by using a lovely 
theorem about non-homogeneous {\em controlled} random walks by Durrett, Kesten, and
Lawler \cite{DKL91}.

\begin{itemize}

\item[$> 0$:]
The gambler has a strategy 
to go bankrupt with probability $> 0$,  
precisely when there exists an
action $A_i$ such that
there is a negative number $n_{i,j} < 0$
in its support (i.e., in the support of the corresponding
finite-support distribution on the integers).  
If such an action $A_i$ exists, then clearly playing
action $A_i$ repeatedly yields a non-zero probability of eventually 
going bankrupt.
If no such action exists, then the gambler's wealth never decreases
and thus he/she never goes bankrupt, no matter what it does.

\item[$=1$:]   We wish to know whether the gambler has a strategy
with which it will go bankrupt with probability 1.  (Never
mind that the gambler
would be stupid to do this.) 

Note that, by the reduction to OC-MDPs described above, this 
case is equivalent to whether in the resulting
OC-MDP the controller has a strategy to terminate
(i.e., hit counter value $0$) in {\em any} state, with probability 1.
Note that this is the {\em non-selective} termination condition (NT).
Thus by Theorem \ref{thm-NT}, if the supremum probability, over
all strategies, of terminating is $1$, then  
there is in fact a {\em counter-oblivious memoryless} (CMD)
optimal strategy, $\sigma$, for terminating with probability 1.
But note that there is only one controlled state in the OC-MDP (the state $s$),
from which the controller chooses one of the actions
$A_1, \ldots, A_k$.
Thus, the CMD strategy $\sigma$ amounts to always choosing the same
action, $A_i$.
Translating this strategy back to the solvency game, 
if the supremum probability of bankruptcy is $1$, then there is
an optimal action $A_i$ that the gambler should choose repeatedly for ever,
which achieves bankruptcy probability $= 1$.

How do we decide which action does this?  This is simple: 
let the {\em drift}, $E[A_i]$, associated with an action $A_i$ be the expected
change in the counter value if we take action $A_i$ once.
This can clearly be computed easily in polynomial time from the description
of the probability distribution for $A_i$.  

Note that once we fix an action $A_i$ that we will choose forever, 
this basically yields a 1-dimensional homogeneous random walk on
the integers, starting from a positive integer. 
It then follows from a basic results in the theory of random walks and
sums of i.i.d. random variables (see, e.g., \cite{Chung01} 
Theorem 8.2.5 and Theorem 8.3.4) that, fixing action $A_i$, 
the resulting random walk (starting with a positive wealth) will hit 
wealth $0$ (bankruptcy) with probability 1 if and only if both
of the following conditions hold: (1) $E[A_i] \leq 0$ (i.e., the
drift is not positive, and (2) $A_i$ has some negative value $n_{i,j} < 0$
in its support.

We can of course check these conditions individually for each action $A_i$,
and we answer yes precisely if some action satisfies these conditions.

\item[$= 0$:]
Is there a strategy for the gambler to not go bankrupt with probability $1$? 
Clearly, this is the case if and only if there exists an action $A_i$
which {\em does not} have a negative number $n_{i,j} <0$ in its support.
It is trivial to check this.

\item[$< 1$:]  
Finally, we come to the most interesting and difficult case:
is there a strategy for the gambler to go bankrupt with probability $< 1$,
i.e., to not go bankrupt with positive probability?

Note that:

\begin{enumerate}
\item If there exists an action $A_i$ which 
does not have  a negative integer $n_{i,j} < 0$ in its support,
then playing that action repeatedly suffices to not go bankrupt
(in fact to not go bankrupt with probability 1).  

\item If there exists an action $A_i$ 
such that
$E[A_i] > 0$  (i.e., whose {\em drift} is positive), then
again by basic facts about random walks and sums of i.i.d.
random variables (again, see, e.g., Theorems 8.2.5 and 8.3.4 of
\cite{Chung01}), starting with any positive wealth, with positive
probability the wealth will never hit $0$.
\end{enumerate}

Clearly, both conditions (1.) and (2.) 
can be checked easily in polynomial time.

Is there any other possible way for the gambler to not go bankrupt
with positive probability, perhaps by using 
some combination
of different actions as its strategy?  
We shall now see that this is not possible.  
If no action satisfies either of the above two conditions, then
there is no strategy at all for the gambler to not go bankrupt with
positive probability.

This follows for a lovely (and quite non-trivial to prove) result
due to Durrett, Kesten and
Lawler \cite{DKL91} about non-homogeneous {\em controlled} random walks
(or, as they put it, 
about when one can and cannot {\em ``make money from fair games''}).
Specifically, Theorem 1 of \cite{DKL91} says the following:  suppose 
a gambler gets to choose a sequence  
$X_1, X_2, X_3, \ldots$ of {\em independent} 
random variables whose range is over the reals,  
such that 
the $X_i$'s, although not necessarily identically distributed,
do have the property that they are
only {\em finitely inhomogeneous}, 
meaning that 
there exists a finite family of probability distributions
$\F = \{F_1, \ldots, F_k\}$ over the reals, such that
for all $i \in \Nset$, 
the distribution of $X_i$ comes from the family $\F$.
Suppose, furthermore, that
every distribution in $\F$ has mean $0$, i.e., $E[X_i] =0$, for all $i$, and
has {\em finite non-zero variance}, i.e., $0 < Var[X_i] < \infty$,
for all $i$. 
Let $S_n = \sum^n_{i=1} X_i$, for $n \in \Nset$.
The gambler's stategy can be {\em adapted}, meaning its choice
of distribution for $X_i$ can depend on the outcomes from $X_1,\ldots,X_{i-1}$.

Theorem 1 of \cite{DKL91} says that
as long as these conditions hold, the sequence of random
variables $S_n$ is {\em recurrent}, meaning there
is some $0< L < \infty$ such
that $\Prob(S_n \in [-L,L] \ i.o.) = 1$, or in other words, such that
the probability that $S_n \in [-L, L]$ infinitely often
(i.e., for infinitely many $n$) is 1.\footnote{Incidentally, 
in \cite{DKL91} they also note that without the condition that $Var[X_i] < \infty$,
there are simple examples where   $S_n \rightarrow \infty$ almost surely.
In other words, without such conditions on higher moments, one can indeed {\em make
money from fair games.}}
Note that this also means that for an fixed value $D < 0$, 
with probability 1 the sequence $S_n$
will eventually hit a value $\leq D$.
(This is because it will have infinitely many ``shots'' at hitting
 a value $\leq D$ from a starting  point inside the interval $[-L,L]$,
 and each such shot has a positive probability which is bounded
 away from $0$ by a positive $\epsilon >0$.  This later fact holds
 because there are only finitely many distributions to choose from,
 and each distribution is non-trivial because it has non-zero variance.)

Let us see now why this implies that the {\em only} conditions
under which the gambler has a strategy not to go bankrupt 
with positive probability are when either one of conditions ($1.$)
or ($2.$) above hold.

Consider the set of actions $A_1, \ldots, A_k$.  
Suppose neither condition ($1.$) nor ($2.$) holds for any of these
actions.  Thus, each action $A_i$ has some negative 
integer $n_{i,j} < 0$ in its support, and furthermore no action
$A_i$ has positive drift, i.e., for all actions $A_i$, $E[A_i] \leq 0$.

Let us first assume that all actions have drift $0$, i.e., for all
$i$, $E[A_i] = 0$.  In this case, since each action has a negative
integer in its support, clearly $Var[A_i] > 0$.  
Furthermore, for every $i$ the distribution of $A_i$ has only finite
support, clearly $Var[A_i] < \infty$.
Thus we are in exactly the situation of Theorem 1 of \cite{DKL91},
and consequently we know that regardless of what wealth $D$ we start with,
with probability $1$ the wealth will eventually hit a value $\leq 0$.

What if there are some actions $A_i$ for which $E[A_i] < 0$?
Well, intuitively, this can only favor the probability of bankruptcy.
More formally, we can do as follows:  
for every action $A_i$ with $E[A_i] < 0$, 
obtain a new random
variable $A'_i$ from $A_i$  by letting $A'_i =  A_i - E[A_i]$.
Clearly,  $E[A'_i] = 0$.  Furthermore, $0 < Var[A'_i] < \infty$,
because the same holds for $A_i$.
Thus, for these revised random variables, again, the condition
holds that starting from any positive wealth the gambler eventually
goes bankrupt with probability 1, regardless of the strategy.
But sums of these revised random variables are always just rightward
translations of sums of the original set of random variables.
So if we go bankrupt with probability 1 with the revised random
variables, then we would also go bankrupt with probability 1 
with the original random
variables. This completes the proof.\\  Thus  
checking cases ($1.$) and ($2.$) for each action 
yields a correct polynomial time algorithm for determining
whether there is a strategy for the gambler to not go bankrupt with
positive probability.
\end{itemize}
\qed
\end{proof}